\pdfoutput=1
\documentclass[11pt]{article}

\usepackage{arxiv}
\usepackage[utf8]{inputenc} 
\usepackage[T1]{fontenc}    
\usepackage{hyperref}       
\usepackage{url}            
\usepackage{booktabs}       
\usepackage{amsfonts}       
\usepackage{nicefrac}       
\usepackage{microtype}      
\usepackage{nomencl}
\usepackage{amssymb}
\usepackage{amsmath}
\usepackage{subcaption}
\usepackage[normalem]{ulem}
\hypersetup{colorlinks,urlcolor=blue}
\usepackage[bbgreekl]{mathbbol}
\usepackage{etoolbox}
\usepackage{mathtools}
\usepackage{algpseudocode}
\usepackage{color}
\usepackage{placeins}
\usepackage[style=numeric-comp,backend=biber, giveninits=true, sorting=none, maxbibnames=9]{biblatex}
\DeclareSymbolFontAlphabet{\mathbb}{AMSb}
\DeclareSymbolFontAlphabet{\mathbbl}{bbold}

\newcommand{\krateeq}{\mathrm{k}_{\ao}^{\text{eq}}}

\newcommand{\kratem}{\mathrm{k}_{\am}^{\text{eq}}}

\newcommand{\as}{\alpha_{\text{s}}}

\newcommand{\ao}{\alpha}

\newcommand{\Xalpha}{\mathrm{X}_{\alpha}}

\newcommand{\Xa}{\mathrm{X}_{\alpha_{\text{s}}}}

\newcommand{\Xm}{\mathrm{X}_{\am}}

\newcommand{\Xw}{\mathrm{X}_{\aw}}

\newcommand{\Xgb}{\mathrm{X}_{\agb}}

\newcommand{\Xb}{\mathrm{X}_{\beta}}

\newcommand{\Xs}{\mathrm{X}_{\text{sol}}}

\newcommand{\Xl}{\mathrm{X}_{\text{liq}}}

\newcommand{\aw}{\alpha_{\text{w}}}

\newcommand{\agb}{\alpha_{\text{gb}}}

\newcommand{\am}{\alpha_{\text{m}}}

\newcommand{\dXa}{\dot{\mathrm{X}}_{\alpha_{\text{s}}}}

\newcommand{\dXm}{\dot{\mathrm{X}}_{\am}}

\newcommand{\dXb}{\dot{\mathrm{X}}_{\beta}}
\newcommand{\dXs}{\dot{\mathrm{X}}_{\text{sol}}}

\newcommand{\dXbtoa}{\dot{\mathrm{X}}_{\beta\rightarrow\as}}
\newcommand{\dXmtoa}{\dot{\mathrm{X}}_{\am\rightarrow\as}}
\newcommand{\dXmtob}{\dot{\mathrm{X}}_{\am\rightarrow\beta}}
\newcommand{\dXatob}{\dot{\mathrm{X}}_{\as\rightarrow\beta}}
\newcommand{\dXbtoam}{\dot{\mathrm{X}}_{\beta\rightarrow\am}}

\newcommand{\Xbeq}{\mathrm{X}_{\beta}^{\text{eq}}}
\newcommand{\Xaeq}{\mathrm{X}_{\alpha}^{\text{eq}}}

\newcommand{\Xmeq}{\mathrm{X}_{\am}^{\text{eq}}}
\newcommand{\Xmeqo}{\mathrm{X}_{\am,0}^{\text{eq}}}

\newcommand{\Tms}{T_{\alpha_{\text{m}},\text{sta}}}
\newcommand{\dT}{\dot{T}}
\newcommand{\dTm}{\dot{T}_{\alpha_{\text{m}}, \text{min}}}
\newcommand{\Ts}{T_{\text{sol}}}
\newcommand{\Tl}{T_{\text{liq}}}
\newcommand{\Tig}{T_{\text{ig}}}

\newcommand{\Tbs}{T_{\alpha_s,\text{end}}}
\newcommand{\Tbe}{T_{\alpha_s,\text{sta}}}

\DeclarePairedDelimiterX{\infdivx}[2]{}{}{%
  #1\;\big\|\;#2%
}

\usepackage{amsthm}
\newcommand\newsubcap[1]{\phantomcaption\caption*{\figurename~\thefigure\thesubfigure: #1}}
\makeatletter
\newtheoremstyle{indented}
  {3pt}
  {3pt}
  {\addtolength{\@totalleftmargin}{1.5em}
   \addtolength{\linewidth}{-1.5em}
   \parshape 1 1.5em \linewidth}
  {}
  {\bfseries}
  {.}
  {.5em}
  {}
\makeatother
\theoremstyle{indented}

\newtheorem*{remark}{Remark}
\newlength{\commentindent}
\makeatletter

\LetLtxMacro{\oldalgorithmic}{\algorithmic}
\renewcommand{\algorithmic}[1][0]{%
  \oldalgorithmic[#1]%
}
\makeatother 
\allowdisplaybreaks 

\title{A Novel Physics-Based and Data-Supported Microstructure Model for Part-Scale Simulation of Laser Powder Bed Fusion of Ti-6Al-4V}

\author{
  Jonas Nitzler*\\
  Institute for Computational Mechanics\\
  Technical University of Munich\\
  D-85748 Garching b. München\\
  \texttt{nitzler@lnm.mw.tum.de} \\
  \And
    Christoph Meier*\\
 Institute for Computational Mechanics\\
  Technical University of Munich\\
  D-85748 Garching b. München\\
  \texttt{meier@lnm.mw.tum.de} \\
   \And 
   Kei W. M\"uller \\
 Institute for Computational Mechanics\\
  Technical University of Munich\\
  D-85748 Garching b. München\\
  \texttt{mueller@lnm.mw.tum.de} \\
  \And
Wolfgang A. Wall\\
Institute for Computational Mechanics\\
  Technical University of Munich\\
  D-85748 Garching b. München\\
  \texttt{wall@lnm.mw.tum.de} \\
   \And
 Neil E. Hodge \\
 Lawrence Livermore National Laboratory\\
 Livermore, CA 94550-9234 \\
  \texttt{hodge3@llnl.gov} \\
}

\begin{document}

\maketitle

\begin{center}
*shared first authorship
\end{center}

\begin{abstract}
The elasto-plastic material behavior, material strength and failure modes of metals fabricated by additive manufacturing technologies are significantly determined by the underlying process-specific microstructure evolution. In this work a novel physics-based and data-supported phenomenological microstructure model for Ti-6Al-4V is proposed that is suitable for the part-scale simulation of selective laser melting processes. The model predicts spatially homogenized phase fractions of the most relevant microstructural species, namely the stable $\beta$-phase, the stable $\as$-phase as well as the metastable Martensite $\am$-phase, in a physically consistent manner. In particular, the modeled microstructure evolution, in form of diffusion-based and non-diffusional transformations, is a pure consequence of energy and mobility competitions among the different species, without the need for heuristic transformation criteria as often applied in existing models. The mathematically consistent formulation of the evolution equations in rate form renders the model suitable for the practically relevant scenario of temperature- or time-dependent diffusion coefficients, arbitrary temperature profiles, and multiple coexisting phases. Due to its physically motivated foundation, the proposed model requires only a minimal number of free parameters, which are determined in an inverse identification process considering a broad experimental data basis in form of time-temperature transformation diagrams. Subsequently, the predictive ability of the model is demonstrated by means of continuous cooling transformation diagrams, showing that experimentally observed characteristics such as critical cooling rates emerge naturally from the proposed microstructure model, instead of being enforced as heuristic transformation criteria. Eventually, the proposed model is exploited to predict the microstructure evolution for a realistic selective laser melting application scenario and for the cooling/quenching process of a Ti-6Al-4V cube of practically relevant size. Numerical results confirm experimental observations that Martensite is the dominating microstructure species in regimes of high cooling rates, e.g. due to highly localized heat sources or in near-surface domains, while a proper manipulation of the temperature field, e.g., by preheating the base-plate in selective laser melting, can suppress the formation of this metastable phase.
\end{abstract}


\section{Introduction}
\label{sec: introduction}
Additive manufacturing has become an enabler for next-generation mechanical designs with applications ranging from complex geometries for patient-specific implants to custom lightweight structures for the aerospace industry. Especially metal selective laser melting (SLM) has gained broad interest due to its high quality and flexibility in the manufacturing process of load-bearing structures. Still, the reliable certification of such parts is an open research field not least because of a multitude of complex phenomena requiring the modeling of interactions between several physical domains on macro-, meso- and micro-scale~\cite{meier2017}. 

A significant impact on elastoplastic material behavior, failure modes and material strength is imposed by the evolving microstructural composition during the SLM process~\cite{Ronda.1996,Furrer.2010,Kelly.Diss,Yang.2016}. Modeling the microstructure evolution in selective laser melting is thus an important aspect for more reliable and accurate process simulations and a crucial step towards certifiable computer-based analysis for SLM parts.

In \cite{Furrer.2010, Murgau.Diss} microstructure models are divided into the categories statistical \cite{Malinov.2000,Mishra.2004,Yang.2000,Ding.2004,Grujicic.2001,Kar.2006,reddy2008prediction,rai2016coupled,koepf2019numerical,zhang2013probabilistic,nie2014numerical}, phenomenological~\cite{Kelly.Diss,Murgau.Diss, Murgau.C, Murgau.D,Ronda.1996, Malinov.2001.Differential,Luetjering.1998,Fan.2005,Crespo.2011,Porter.2009,Furrer.2010,Grong.2002,salsi2018modeling,lindgren2016simulation} and phase-field~\cite{Katzarov.2002,Radhakrishnan.2016,Gong.2014,Chen.2004,gong2015phase} models. Statistical models are either data-driven and infer statements of coarse-grained trends from experiments or apply local stochastic transformation rules and neighborhood dependencies, which might be based on physical principles. This category includes in the context of this work also data-based surrogates and machine learning approaches as well as Monte-Carlo simulations and (stochastic) cellular automaton approaches. Without physical foundation, the reliability and predictive ability of purely data-driven approaches is rather limited, especially in the case of very scarce and expensive experimental data (e.g., dynamic microstructure characteristics in the high-temperature regime) or if the available data does not contain certain physical phenomena at all, which might result in high generalization errors. In case the simulation is based on stochastic rules (e.g., Monte-Carlo simulations) one encounters often challenges in terms of computationally demands as a reliable response statistic requires a large number of simulation runs. Furthermore, a consistent conservation of global and local physical properties for the individual simulation runs remains an open challenge. Stochastic properties might additionally be space and time dependent or functionally dependent on further physical properties. The inference of suitable and generalizable parameterizations of these stochastic properties is especially problematic in the case of limited experimental data.

On the other end of the spectrum of available models, phase-field approaches offer the greatest insight into microstuctural evolution and provide a detailed resolution of the underlying physics-based phenomena of crystal formation and dissolution, such as crystal boundaries and lamellae orientation. However, the resolution of length scales below the size of single crystals comes at a considerable computational cost, 
which hampers their application for part-scale simulations.

A preferable cost-benefit ratio can be found in the category of phenomenological modeling approaches. Here, microstructure evolution is described in a spatially homogenized (macroscale) continuum sense by physically motivated, phenomenological phase fraction evolution laws that can be solved at negligible extra cost as compared to standard thermal (or thermo-mechanical) process simulations. In this work we will propose a novel physics-based and data-supported phenomenological microstructure model for Ti-6Al-4V that is suitable for the part-scale simulation of selective laser melting processes. We present several original contributions, compared to existing approaches of this type. Compared to existing approaches of this type several original contributions, both in terms of physical and mathematical consistency but also in terms of the underlying data basis, can be identified. 

From a physical point of view, the phase fraction evolution equations proposed in this work are solely motivated by energy considerations, i.e. deviations from thermodynamic equilibrium configurations are considered as driving forces for diffusion-based and non-diffusional transformations. Thus, the evolution of the most relevant microstructural phases, namely the $\beta$-phase, the stable $\as$-phase as well as the metastable martensitic $\am$-phase, is purely driven by an energetic competition and the temperature-dependent mobility of these different specifies. This is in strong contrast to existing approaches, where e.g. the formation of meta-stable phases is triggered by heuristic rules for critical cooling rates, which are taken from experimental observations and explicitly prescribed in the model to match the former. In the present approach, however, there is no need to \textit{prescribe} such critical cooling rates as \textit{criterion} for phase formation. Instead, phase formation is a pure consequence of the underlying energy and mobility competition. Critical cooling rates can be \textit{predicted} as a \textit{result} of the modeling approach, and show very good agreement to experimental observations. 

From a mathematical point of view, the diffusion-based transformations are described in a consistent manner by evolution equations in differential form, i.e. ordinary differential equations that are numerically integrated in time, which renders the model suitable for the practically relevant case of solid state transformations involving temperature or time-dependent diffusion coefficients, arbitrary temperature profiles, and multiple coexisting phases. Again, this is in contrast to existing approaches modeling the phase evolution with Johnson-Mehl-Avrami-Kolmogorov (JMAK) equations. In fact, JMAK equations can be identified as analytic solutions for differential equations of the aforementioned type, which are, however, not valid anymore in the considered case of non-constant (temperature-dependent) parameters.

From a data science point of view, unknown parameters in existing modeling approaches are typically \textit{calibrated} on the basis of single experiment data. As a consequence, this single experiment can then be represented with very good agreement while an extrapolation of the calibration data, i.e. a truly predictive ability, is only possible within very narrow bounds. In the present approach, a broad basis of experimental data in form of time-temperature transformation (TTT) diagrams is considered for inverse parameter identification of the (small number of) unknown model parameters. Moreover, the predictive ability of the identified microstructure model is verified on an independent data set in form of continuous-cooling transformation (CCT) diagrams, showing very good agreement in the characteristics (e.g. critical cooling rates) of numerically predicted and experimentally measured data sets. Since experimental CCT data is very limited (to only a few discrete cooling curves), the prediction of these diagrams by numerical simulation is not only relevant for model verification. In fact, the proposed microstructure model allowed for the first time to predict CCT data of Ti-6Al-4V for such a broad and highly resolved range of cooling rates, thus providing an important data basis for other researchers in this field. Eventually, the proposed model is exploited to predict the microstructure evolution for a realistic SLM application scenario (employing a state-of-the-art macroscale SLM model) and for the cooling/quenching process of a Ti-6Al-4V cube with practically relevant size (side length $10\ cm$).

The structure of the paper is as follows: Section~\ref{sec: crystallography} briefly presents the relevant basics of Ti-6Al-4V crystallography, the basic assumptions and derivation of the proposed microstructure evolution laws and finally the temporal discretization and implementation of the model in form of a specific numerical algorithm. Section~\ref{sec: calibration} depicts the data-supported inverse parameter identification on the basis of TTT diagrams and model verification in form of CCT diagrams. In Section~\ref{sec: part_scale_demonstration}, first the basics of a thermo-mechanical finite element model employed for the subsequent part-scale simulations are presented. Then, in Sections~\ref{sec: part_scale_demonstration} and~\ref{sec: scaling_effects} applicability of the proposed microstructure model to part scale simulations is demonstrated by means of two practically relevant examples, a realistic SLM application scenario as well as the cooling/quenching process of a Ti-6Al-4V cube.

\section{Derivation of a novel microstructure model for Ti-6Al-4V}
\label{sec: crystallography}
In the following, we will derive a model for the microstructure evolution in Ti-6Al-4V in terms of volume-averaged phase fractions. First, we introduce fundamental concepts and outline our basic assumptions in Section \ref{sec:fundamentals}. Afterwards, equilibrium and pseudo-equilibrium compositions of the microstructure states are presented in Section \ref{sec:equilibrium_states} as a basis for the subsequent concepts for arbitrary microstructure changes in Section \ref{sec:crystallography_evolution}. The model will be presented in a continuous and discretized formulation. The latter is then used in the numerical demonstrations.

\subsection{Crystallographic fundamentals and basic assumptions}
\label{sec:fundamentals}

\begin{figure}[htbp]
 \centering
 \includegraphics[scale=0.36]{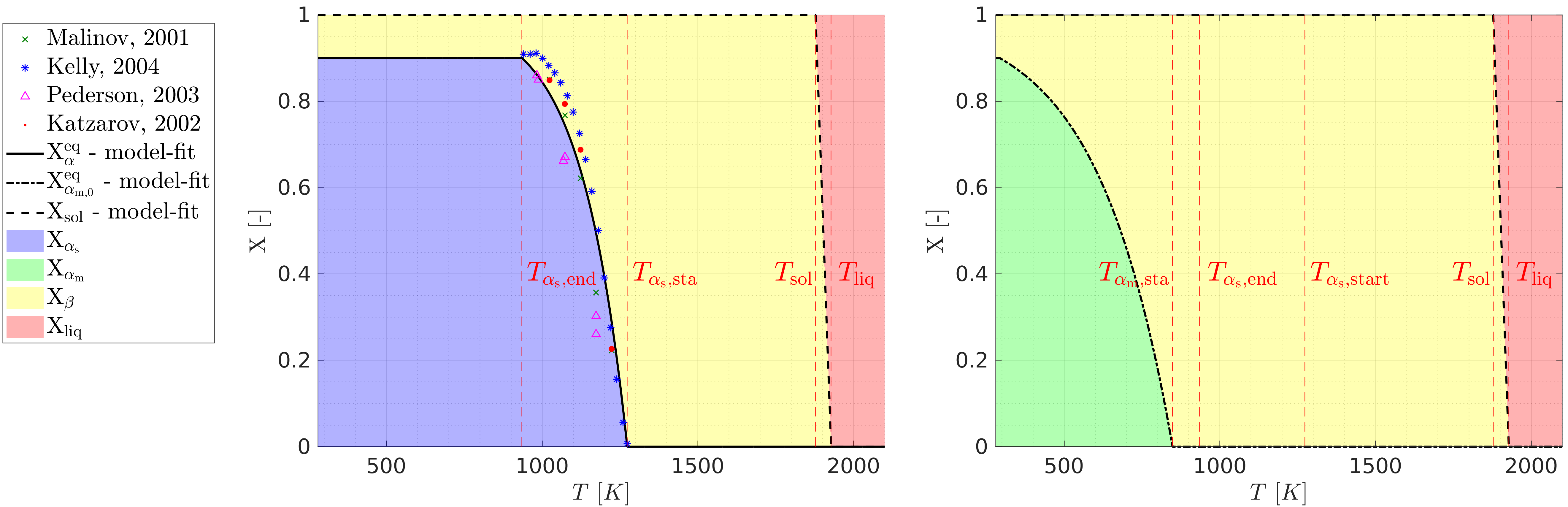}
  \caption{Left: Equilibrium compositions~$\Xa$,~$\Xb$ and~$\Xl$ resulting from slow cooling rates ${|\dT|\! \ll \!|\dTm|}$, such that $\Xm=0$. The experimental data is based on Katzarov \cite{Katzarov.2002}, Kelly \cite{Kelly.Diss}, Malinov \cite{Malinov.2001.Differential} and Pederson \cite{Pederson.2003}. Right: Metastable (pseudo-)equilibrium composition due to Martensite transformation for ${|\dT|\! > \!|\dTm|}$, such that $\Xa=0$.} 
  \label{fig: equilibrium}
 \end{figure}

The aspects of microstructure evolution in Ti-6Al-4V alloys as considered in this work are assumed to be determined by the current microstructural state, the current temperature $T$ as well as the temperature rate $\dT$, i.e. its temporal derivative. An overview of characteristic temperatures are given in Table~\ref{tab: CharacteristicTemperatures}.
\begin{table}[htbp]
		\caption{Characteristic temperatures deployed in the microstructure model}
		\label{tab: CharacteristicTemperatures}
		\centering
		\renewcommand{\arraystretch}{1.5}
		\begin{tabular}{ccccccc}
			\toprule
			$T_{\alpha_m,end}$ [K]&$\Tms$ [K]&$\Tbs$ [K]&$\Tbe$ [K]&$\Ts$ [K]&$\Tl$ [K]\\
			\midrule
			       293      &    848      &     935      &   1273      &    1878     & 1928\\
			    -&\cite{Ahmed.1998,Murgau.Diss,Kelly.Diss}&-&\cite{Kelly.Diss,Fan.2005}&\cite{Elmer.2004}&\cite{Elmer.2004}\\
			\bottomrule
		\end{tabular}
\end{table}
When cooling down the alloy from the molten state, solidification takes place between liquidus temperature~$\Tl$ and solidus temperature~$\Ts$. The co-existent liquid and solid phase fractions in this temperature interval shall be denoted as $\Xl$ and $\Xs=1-\Xl$. Below the solidus temperature~$\Ts$ the microstructure of Ti-6Al-4V is characterized by body-centered-cubic {(bcc)~$\beta$-crystals} and hexagonal-closed-packed {(hcp)~$\alpha$-crystals}. First,~$\beta$-crystals will grow in direction of the maximum temperature gradient for~$\Ts<T\le\Tl$ \cite{Kelly.Diss}. Depending on the prevalent cooling conditions, the~$\alpha$-phase can be further subdivided into stable~$\as$ and metastable~$\am$-phases (Martensite). For sufficiently slow cooling rates~$|\dT| \ll |\dTm|$ the microstructure evolution can follow the thermodynamic equilibrium, i.e. stable~$\as$ nucleates into prior~$\beta$-grains. This diffusion-driven transformation between alpha-transus start temperature $\Tbe$ and alpha-transus end temperature $\Tbs$ results in a temperature-dependent equilibrium composition~$\Xaeq(T)$ (see Figure~\ref{fig: equilibrium}, left) characterized by $90\%$ $\as$- and $10\%$ $\beta$-phase, i.e. phase fractions $\Xa=0.9$ and $\Xb=0.1$, for temperatures below $\Tbs$~\cite{Malinov.2001.Differential, Kelly.Diss, Pederson.2003, Katzarov.2002}. 

Under faster cooling conditions, the formation rates of the stable $\as$-phase, which are thermally activated and limited by the diffusion-driven nature of this transformation process, cannot follow the equilibrium composition~$\Xaeq(T)$ anymore such that $\beta$-phase fractions higher than $10\%$ remain below $\Tbs$. At temperatures below the Martensite-start-temperature~$\Tms$, the metastable Martensite-phase $\am$ becomes energetically more favorable than the excess (transformation-suppressed) $\beta$-phase fraction beyond $10\%$. Under such conditions the~$\beta$-crystals collapse almost instantaneously into metastable Martensite following a temperature-dependent (pseudo-) equilibrium composition~$\Xmeq(T)$~\cite{Xu.2015.martensite,Ahmed.1998,Fan.2005,Murgau.Diss,Kelly.Diss}. The critical cooling rate $\dTm$ is defined as the cooling rate above which the formation of stable $\as$ is completely suppressed (up to the precision of measurements). The resulting microstructure consists exclusively of $\beta$- and $\am$-phase fractions. The temperature-dependent Martensite phase fraction is denoted as~$\Xmeqo(T)$ for this extreme case. It is typically reported that the Martensite-end-temperature $T_{\am,\text{end}}$, i.e. the temperature when the Martensite formation is finished, is reached at room temperature $T_\infty$ going along with a maximal Martensite phase fraction of $\Xm=0.9$ in this extreme case (see Figure~\ref{fig: equilibrium}, right).

In addition to these most essential microstructure features, which will be considered in the present work, some additional microstructural distinctions are often made in the literature \cite{Kelly.Diss,Kelly.article,Murgau.Diss}. For example, the stable alpha phase $\as$ can be further classified into its morphologies grain-boundary-$\agb$ and Widmanst\"atten-$\aw$ phase, with ${\Xa=\Xgb + \Xw}$. During cooling, grain-boundary-$\agb$ forms first between the individual~$\beta$-crystals. Upon passing the intergranular nucleation temperature~$\Tig$, lamellae-shaped Widmanst\"atten-$\aw$ grow into the prior~$\beta$-crystal starting from the grain boundary. The appearance of such~$\alpha$-morphologies can have different manifestations such as colony- and basketweave-$\as$ or equiaxed-$\as$-grains \cite{Kelly.Diss}. 

The microstructure model proposed in this work will only consider the accumulated $\as$ phase without further distinguishing $\agb$ and $\aw$ morphologies. Given their similar mechanical properties \cite{Lee.1990, Lee.1991}, this approach seems to be justified, as the proposed microstructure model shall ultimately be employed to inform homogenized, macroscale constitutive laws for the part-scale simulation of metal powder bed fusion additive manufacturing (PBFAM) processes. In a similar fashion, the so-called \textit{massive transformation} is often observed in the range of cooling rates that are sufficiently high but still below $\dTm$, which leads to microstructural properties laying between those of pure $\as$ and pure $\am$. For similar reasons as argued above, this microstructural species is not explicitly resolved by a separate phase variable in the present work. Instead, the effective mechanical properties of this intermediate phases are captured implicitly by the co-existence of $\as$ and $\am$ phase resulting from the present model at these cooling conditions. In addition and in summary, the following basic assumptions are made for the proposed modeling approach:
\begin{enumerate}
\item The microstructure is described in terms of (volume-averaged) phase fractions, i.e. no explicit resolution of grains and grain boundaries.
\item Only the most important phase species $\beta$, $\as$ and $\am$ are considered.
\item The Martensite-start- and Martensite-end-temperature are considered as constant, i.e. independent of the current microstructure configuration.
\item Currently, no information about (volume-averaged) grain sizes, morphologies and orientations is provided by the model.
\item The influence of the mechanical stress state, microstructural imperfections (e.g. dislocations) as well as further morphologies is not considered.
\end{enumerate}

Partly, these assumptions are motivated by a lack of corresponding experimental data. In our ongoing research work, we intend to address several of these limitations.

\begin{remark}[Cooling rates during quenching]
Note that the cooling rates during quenching experiments are not temporally constant in general. Thus, the critical cooling rate $\dTm$ measured in experiments is usually the cooling rate at one defined point in time, typically defined at a high temperature value such that the measured cooling rate is (close to) the maximal cooling rate reached during the quenching process. In such a manner, a value of $\dTm=410\ K/s$ has been reported in the literature for Ti-6Al-4V~\cite{Ahmed.1998} but was interpreted in several contributions \cite{Murgau.Diss, Fan.2005, Irwin.2017} as a fixed constraint for Martensite transformation.
\end{remark}
 
\subsection{Equilibrium and pseudo-equilibrium compositions}
\label{sec:equilibrium_states}

In this Section the temperature-dependent, thermodynamic equilibrium and pseudo-equilibrium compositions of the $\beta$, $\as$ and $\am$ are described in a quantitative manner. These phase fractions $\mathrm{X}_i \in [0;1]$ have to fulfill the following continuity constraints:
\begin{subequations}
\label{eqn: continuity_constraints}
\begin{align}
    \Xs+\Xl&=1,\label{eqn: continuity_constraints1}\\
    \Xalpha+\Xb &=\Xs,\label{eqn: continuity_constraints2}\\
    \Xa + \Xm &= \Xalpha,\label{eqn: continuity_constraints3}
\end{align}
\end{subequations}
For simplicity, the solidification process between liquidus temperature~$\Tl$ and solidus temperature~$\Ts$ is modeled via a linear temperature-dependence of the solid phase fraction $\Xs$:
\begin{equation}
\label{eqn: solid_phase}
\Xs=
\begin{cases}
1&\text{for }\ T\le \Ts,\\
1 - \frac{1}{\Tl-\Ts}\cdot (T-\Ts)& \text{for }\ \Ts<T<\Tl,\\
0&\text{for }\ T\ge\Tl.
\end{cases}
\end{equation}
Moreover, we follow the standard approach to model the temperature-dependent \textit{stable} equilibrium phase fraction~$\Xaeq(T)$, towards which the $\as$-phase tends in the extreme case of very slow cooling rates~$|\dT| \ll |\dTm|$, on the basis of an exponential Koistinen-Marburger law (~\cite{koistinen}; see black solid line in Figure~\ref{fig: equilibrium}): 
\begin{align}
\label{eqn: mean_approx}
\begin{split}
 \Xaeq(T)&=
\begin{cases}
0.9&\text{for }T<\Tbs,\\
1-\exp\left[-\krateeq\cdot\left(\Tbe-T\right)\right]&\text{for }\Tbs\le T\le \Tbe,\\
0&\text{for }T>\Tbe.
\end{cases}
\end{split}
\end{align}
While the alpha-transus start temperature $\Tbe\!=\!1273K$ has been taken from the literature~\cite{Kelly.Diss,Fan.2005}, the parameters $\Tbs\!=\!935K$ and $\krateeq\!=\!0.0068 \, K^{-1}$ in~\eqref{eqn: mean_approx} have been determined via least-square fitting based on different experimental measurements as illustrated in Figure~\ref{fig: equilibrium} on the left side. It has to be noted that the equilibrium composition $\Xaeq=f(T)$ in form of a temperature-dependent function $f(T)$ as given in~\eqref{eqn: mean_approx} could alternatively be derived as the stationary point $(\partial \Pi(\Xa,T) / \partial \Xa) |_{(\Xa=\Xaeq)} \dot{=} \, 0 \Leftrightarrow \Xaeq - f(T) \, \dot{=} \, 0$ of a generalized thermodynamic potential $\Pi(\Xa,T)$ containing contributions, e.g., from the Gibb's free energies of the individual phases $\beta$ and $\as$, from phase/grain boundary interface energies or from (transformation-induced) strain energies~\cite{Ronda.1996}. Here, the dependence of the potential $\Pi(\Xa,T)$ on $\Xb$ has been omitted since $\Xb=1-\Xa$ for solid material under equilibrium conditions, i.e. in the absence of Martensite. In the present work, for simplicity, the expression for the equilibrium composition $\Xaeq=f(T)$ has directly been postulated and calibrated on experimental data instead of formulating the individual contributions of a potential, which would involve additional unknown parameters. Still, from a mathematical point of view it shall be noted that the expression $\Xaeq=f(T)$ in~\eqref{eqn: mean_approx} is integrable, i.e., a corresponding potential can be found in general, resulting in beneficial properties not only of the physical model but also of the numerical formulation.\\

Next, we consider the second extreme case of very fast cooling rates $|\dT| \geq |\dTm|$ at which the diffusion-driven formation of $\Xa$ is completely suppressed. For this case, we model the \textit{metastable} Martensite pseudo equilibrium fraction $\Xmeqo(T)$, emerging in the absence of $\as$-phase, based on an exponential law \cite{Gil.1996,Fan.2005,Kelly.Diss,Murgau.Diss,koistinen}: 
\begin{align}
\label{eqn: am_equo}
\begin{split}
 \Xmeqo(T)&=
\begin{cases}
0.9&\text{for }T<T_\infty,\\
1-\exp\left[-\kratem(\Tms-T)\right]&\text{for }T_\infty \le T \le \Tms,\\
0&\text{for }T>\Tms.
\end{cases}
\end{split}
\end{align}
While the value $\Tms\!=\!848K$ has been taken from the literature~\cite{Ahmed.1998,Murgau.Diss,Kelly.Diss}, we choose~$\kratem\!=\!0.00415 \, K^{-1}$ such that~\eqref{eqn: am_equo} yields a maximal Martensite fraction of $\Xmeq(T_\infty)\!=\!0.9$ at room temperature, which is in agreement to corresponding experimental observations~\cite{Fan.2005} (see Figure~\ref{fig: equilibrium} on the right).\\

Finally, we want to consider the most general case of cooling rates that are too fast to complete the diffusion-driven formation of the stable $\as$ phase before reaching the Martensite start temperature $\Tms$ but still below the critical rate  $|\dTm|$, i.e. a certain amount of stable $\as$ phase has still been formed and consequently a Martensite phase fraction below $90\%$ is expected at room temperature. For this case, we postulate an effective pseudo equilibrium phase fraction $\Xmeq(T)$ for the $\am$ phase that accounts for the reduced amount of transformable $\beta$-phase at presence of a given phase fraction $\Xa$ of the stable $\as$ phase according to
\begin{align}
\label{eqn: am_eq_general}
 \Xmeq(T)= \Xmeqo(T) \cdot \frac{(0.9-\Xa)}{0.9}.
\end{align}
It can easily be verified that~\eqref{eqn: am_eq_general} fulfills the important relation $ \Xmeq(T)\!+\!\Xa\!<\!0.9$ for arbitrary values of the current temperature $T$ and $\as$-phase fraction $\Xa$. This means, for any given value $\Xa$, an instantaneous Martensite formation according to $\Xmeq(T)$ will never result in a total $\alpha$-phase fraction $\Xalpha\!=\! \Xm\!+\!\Xa$ that exceeds the corresponding equilibrium composition $\Xaeq$ (which takes on a value of $\Xaeq \!=\!0.9$ in the relevant temperature range below $\Tbs$). In the extreme case that the maximal $\as$-phase fraction of $90\%$ has already been formed before reaching the Martensite start temperature $\Tms$, Equation~\eqref{eqn: am_eq_general} ensures 
that no additional Martensite is created during the ongoing cooling process. Again, the pseudo equilibrium composition $\Xmeq=\tilde{f}(T)$ in form of a temperature-dependent function $\tilde{f}(T)$ as given in~\eqref{eqn: am_eq_general} could alternatively be derived as the stationary point $(\partial \Pi(\Xm,\Xa,T) / \partial \Xm ) |_{(\Xm=\Xmeq)} \dot{=} \, 0 \Leftrightarrow \Xmeq - \tilde{f}(T) \dot{=}  0$ of a generalized thermodynamic potential $\Pi(\Xm,\Xa,T)$, in which the current $\as$ phase fraction $\Xa \neq \Xaeq$ can be considered as a fixed parameter. Thus, $\Xmeq=\tilde{f}(T)$ according to~\eqref{eqn: am_eq_general} is not a global minimum of this generalized potential but rather a local minimum with respect to $\Xm$ under the constraint of a given $\as$ phase fraction $\Xa \neq \Xaeq$. This model seems to be justified given the considerably slower formation rate of the $\as$ phase as compared to the (almost) instantaneous Martensite formation (see also the next section).\\

For a given temperature $ \Tms \geq T \geq T_\infty$ and $\as$-phase fraction $\Xa \leq 0.9$ during a cooling experiment, Equation~\eqref{eqn: am_eq_general} will in general yield a Martensite phase fraction such that $\Xalpha = \Xa + \Xm \leq 0.9$, i.e. the sum of stable and martensitic alpha phase fraction might be smaller than the equilibrium phase fraction $\Xaeq$ according to~\eqref{eqn: mean_approx}. In this case of co-existing $\as$- and $\am$-phase, it is assumed that $\Xaeq$  in~\eqref{eqn: mean_approx} as well as its complement $\Xbeq=1-\Xaeq$ represent the (pseudo-) equilibrium compositions for the total $\alpha$ phase fraction $\Xalpha = \Xa + \Xm$ and for the $\beta$ phase fraction $\Xb=1-\Xalpha$. In other words, the driving force for diffusion-based $\as$-formation, as discussed in the next section, is assumed to result in the following long-term behavior:
\begin{align}
    \label{eqn: alpha_total_equ_new}
    \lim_{t \rightarrow \infty}  \Xalpha = \Xaeq  \quad \Leftrightarrow \quad \lim_{t \rightarrow \infty}  \Xb = \Xbeq \quad \text{with} \quad \Xalpha=\Xa+\Xm, \,\,\, \Xb=1-\Xalpha, \,\,\, \Xbeq=1-\Xaeq.
\end{align}
While at low temperatures, Martensite is energetically more favorable than the $\beta$-phase, which is the driving force for the instantaneous Martensite formation, it is assumed to be less favorable than the $\as$-phase. Therefore, there exists a driving force for a diffusion-based dissolution of Martensite into $\as$-phase, resulting in the following long-term behavior: 
\begin{align}
    \label{eqn: alpha_martensite_equ_new}
    \lim_{t \rightarrow \infty}  \Xm = \bar{\mathrm{X}}_{\am}^{\text{eq}} = 0.
\end{align}
However, as discussed in the next section, the diffusion rates for this thermally activated process drop to (almost) zero at low temperatures such that Martensite is retained as meta-stable phase at room temperature. Thus,~\eqref{eqn: alpha_martensite_equ_new} can only be considered as theoretical limiting case in this low temperature region.

\subsection{Evolution equations}
\label{sec:crystallography_evolution}

Since the melting and solidification process is completely described by~Equation~\eqref{eqn: solid_phase}, this section focuses on solid-state phase transformations for temperatures~$T<\Ts$ below the solidus temperature (i.e. $\Xs=1$). To model the formation and dissolution of the~$\as$-,~$\am$- and~$\beta$-phase, we propose evolution equations in rate form with the following contributions to the total rates, i.e. to the total time derivatives~$\dXa$, $\dXm$ and $\dXb$, of the three phases:
\begin{subequations}
\label{eqn: definition_rates}
\begin{align}
    \label{eqn: definition_rates_alphas}
    \dXa & = \dXbtoa + \dXmtoa - \dXatob,\\
    \label{eqn: definition_rates_alpham}
    \dXm &= \dXbtoam - \dXmtoa - \dXmtob,\\
    \label{eqn: definition_rates_beta}
    \dXb &= \dXatob + \dXmtob - \dXbtoa - \dXbtoam.
\end{align}
\end{subequations}
Here, e.g., $\dXbtoa$ represents the formation rate of $\as$ out of $\beta$ while $\dXatob$ represents the dissolution rate of $\as$ to $\beta$. The meaning of the individual contributions in~\eqref{eqn: definition_rates}, the underlying transformation mechanisms (e.g. instantaneous vs. diffusion-based) as well as the proposed evolution laws will be discussed in the following. Since the formation and dissolution of phases might follow different physical mechanisms in general, we have intentionally distinguished the (positive) rates $\dot{\mathrm{X}}_{x \rightarrow y} \geq 0$ and $\dot{\mathrm{X}}_{y \rightarrow x} \geq 0$ of two arbitrary phases $x$ and $y$ instead of describing both processes via positive and negative values of one shared variable $\dot{\mathrm{X}}_{y \leftrightarrow x}$. It is obvious that~\eqref{eqn: definition_rates} satisfies the continuity Equation~\eqref{eqn: continuity_constraints} for temperatures $T<\Ts$ below the solidus temperature (with $\Xs=1$ and $\dXs=0$), which reads in differential form:
\begin{subequations}
\label{eqn: continuity_solid_differential}
\begin{align}
   \dXa+\dXm+\dXb=0 \quad \text{if} \quad \Xs=1.
\end{align}
\end{subequations}
Thus, the phase fraction $\Xb=1-\Xa-\Xm \, \forall \, T<\Ts$ can be directly calculated from~\eqref{eqn: continuity_constraints} and only the evolution equations for the phases $\as$ and $\am$ will be considered in the numerical algorithm presented in Section~\ref{sec: microstructure_model_numerical}.

\subsubsection{Time-continuous evolution equations in rate form}
\label{sec: microstructure_model_rateform}

\begin{figure}[htbp]
        \centering
        \includegraphics[scale=0.7]{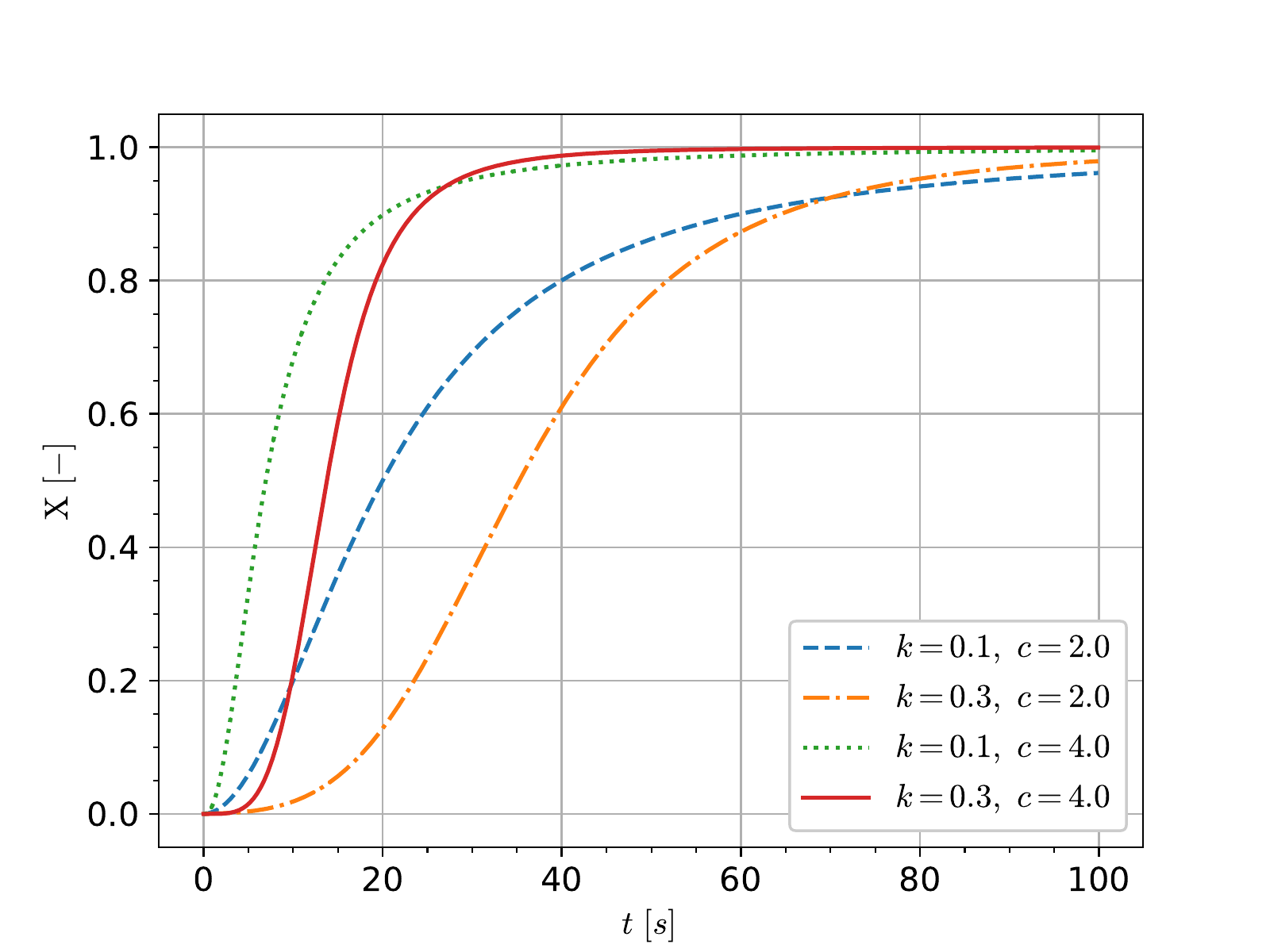}
         \caption{Four exemplary phase evolutions of a nucleation process modelled with Equation \eqref{eqn: alpha_formation_der_new} with initial phase fraction $\mathrm{X}_0=0$, equilibrium $\mathrm{X}^{\text{eq}}=1$ and combinations of $k\in\{0.1,0.3\}$ and $c\in\{2.0,4.0\}$.}
        \label{fig: s_shaped}
\end{figure}

In the following, the individual contributions to the transformation rates in~\eqref{eqn: definition_rates} will be discussed. One of the main assumptions for the following considerations is that $\am \leftrightarrow \beta$ transformations take place on much shorter time scales than $\as \leftrightarrow \beta$ transformations~\cite{Kelly.Diss, Yang.2016, Luetjering.1998}, which allows to consider the former as instantaneous processes while the latter are modeled as (time-delayed) diffusion processes. In a first step, the diffusion-based formation of the stable~$\as$-phase out of the $\beta$-phase is considered~\cite{Kelly.article, Elmer.2004}. Modified logistic differential Equations~\cite{avramov2014generalized} can be considered as suitable model and powerful mathematical tool to model diffusion processes of this type. Based on this methodology, we propose the following model for the diffusion-based transformation $\beta \rightarrow \as$:
\begin{equation}
    \label{eqn: alpha_formation_der_new}
    \dXbtoa = 
    \begin{cases}
        k_{\as}(T) \cdot \left(\Xa \right)^{\frac{c_{\as}-1}{c_{\as}}} \cdot \left(\Xb -\Xbeq \right)^{\frac{c_{\as}+1}{c_{\as}}}& \text{for }\  \Xb > \Xbeq,\\
        0&\text{else}.
    \end{cases}
\end{equation}
Diffusion equations of this type typically consist of three factors: i) The factor $(\Xb\!-\!\Xbeq)$ represents the driving force of the diffusion process in terms of transformable $\beta$ phase (see~\eqref{eqn: alpha_total_equ_new}) and has a decelerating effect on the transformation during the ongoing diffusion process. With the continuity relations $\Xb\!=\!1\!-\!\Xalpha$ and $\Xbeq\!=\!1\!-\!\Xaeq$ this term could alternatively be written as $(\Xaeq\!-\!\Xa)$. ii) The factor with $\Xa$ leads to a transformation rate that increases with increasing amount of created $\as$-phase, i.e., it has an accelerating effect on the transformation rate during the ongoing diffusion process. Physically, this term can be interpreted as representation of the diffusion interface between $\as$- and $\beta$-phase, which increases with increasing size of the $\as$-nuclei (and thus with increasing $\as$-phase fraction). iii) The factor $k_{\as}(T)$ represents the temperature-dependent diffusion rate of this thermally activated process. From a physical point of view, this term considers the temperature-dependent mobility of the diffusing species. When plotting the phase fraction $\Xa$ over time (at constant temperature), the factors i) and ii) together result in the characteristic S-shape of such diffusion-based processes as exemplary depicted in Figure \ref{fig: s_shaped} for four combinations of $k$ and $c$. Depending on the type of diffusion process, different values for the exponent $c_{\as}$ can be derived from the underlying physical mechanisms resulting in more process-specific types of diffusion equations. In its most general form, which is applied in this work, the exponent $c_{\as}$ of the modified logistic differential equation is kept as a free parameter, which allows an optimal inverse identification based on experimental data even for complex diffusion processes~\cite{avramov2014generalized}.

When cooling down ($\dT<0$) the material at temperatures below the Martensite start temperature ($T<\Tms$) and the equilibrium composition of the stable $\as$-phase has not been reached yet ($\Xa < \Xaeq$), an instantaneous Martensite formation out of the (excessive) $\beta$-phase is assumed following the Martensite pseudo equilibrium composition $\Xmeq$ according to~\eqref{eqn: am_eq_general}. Mathematically, this modeling assumption can be expressed by an inequality constraint based on the following Karush-Kuhn-Tucker (KKT) conditions:
\begin{align}
\label{eqn: martensite-formation}
\Xm - \Xmeq  \geq 0 \quad \wedge \quad  \dXbtoam \geq 0 \quad \wedge \quad (\Xm - \Xmeq) \cdot \dXbtoam = 0.
\end{align} 
The constraint $\Xm  \!-\! \Xmeq  \geq 0$ (first inequality in~\eqref{eqn: martensite-formation}) states that $\Xm$ cannot fall below the equilibrium composition $\Xmeq $ since Martensite will be formed instantaneously out of the $\beta$-phase. Here, the formation rate $\dXbtoam $ (second inequality in~\eqref{eqn: martensite-formation}) plays the role of a Lagrange multiplier enforcing the constraint $\Xm \!-\! \Xmeq  = 0$ as long as Martensite is formed. According to the complementary condition (third equation in~\eqref{eqn: martensite-formation}), this formation rate vanishes in case of excessive $\Xm$-phase (i.e. $\dXbtoam=0$ if $\Xm - \Xmeq > 0$). This scenario can occur e.g. during heating of the Martensite material (i.e. $\dot{\mathrm{X}}_{\am}^{\mathrm{eq}}<0$) since the diffusion-based Martensite-dissolution process (see below) cannot follow the decreasing equilibrium composition $\Xmeq$ in an instantaneous manner.

Since the $\am$-phase is energetically less favorable than the $\as$-phase, there is a driving force for the transformation $\am \rightarrow \as$. In contrast to the instantaneous formation of Martensite, the $\am$-dissolution to $\as$ is modeled as (time-delayed) diffusion process~\cite{Murgau.Diss,} according to
\begin{align}
    \label{eqn: alpha_formation_der2_new}
    \dXmtoa \!=\! 
    \begin{cases}
        k_{\as}(T) \!\cdot\! \left(\Xa \right)^{\!\!\!\frac{c_{\as}-1}{c_{\as}}\!} \!\cdot\! \left(\Xm \!-\! \bar{\mathrm{X}}^{\text{eq}}_{\am} \right)^{\!\!\!\frac{c_{\as}+1}{c_{\as}}}\!\!&\text{for }\ \Xm \!>\! \bar{\mathrm{X}}^{\text{eq}}_{\am},\\
        0&\text{else},
    \end{cases}
\end{align}
where $\bar{\mathrm{X}}_{\am}^{\text{eq}}=0$ represents the long-term equilibrium state ($t\rightarrow\infty$) of the Martensite phase (see Equation~\eqref{eqn: alpha_martensite_equ_new}). Equations~\eqref{eqn: alpha_formation_der_new} and~\eqref{eqn: alpha_formation_der2_new} have the same structure and share the same exponent $c_{\as}$ as well as the same diffusion rate $k_{\as}(T)$ since both describe the diffusion-based formation of $\as$-phase, i.e., both result in the same daughter phase. The underlying modeling assumption is that the transformation $\am\rightarrow\as$ can be split according to $\am\rightarrow\beta\rightarrow\as$, i.e., it is assumed that Martensite first dissolves instantaneously into the intermediate phase $\beta$, which afterwards transforms to the $\as$-phase in a diffusion-based $\beta \rightarrow \as$ process similar to~\eqref{eqn: alpha_formation_der_new}. The model for the temperature-dependence of $k_{\as}(T)$ as described below will result in diffusion rates that drop to (almost) zero at low temperatures such that Martensite is retained as meta-stable phase at room temperature, which is in accordance to the corresponding experimental data.

Eventually, also the case of lacking $\beta$-phase (i.e., $\Xb\!<\!\Xbeq$) shall be considered: Due to the nature of the $\beta$-dissolution processes $\dXbtoa$ according to~\eqref{eqn: alpha_formation_der_new} and $\dXbtoam$ according to~\eqref{eqn: martensite-formation}, which only yield contributions as long as $\Xb \!>\! \Xbeq$, this scenario can only arise from $\dot{\mathrm{X}}_{\beta}^{\text{eq}}\!>\!0$, i.e., if $\dot{T} \!>\!0$ and $T \in [\Tbs; \Tbe]$.

Experimental data by Elmer et al. \cite{Elmer.2004} strongly supports a diffusional behavior of $\beta$-phase build up, respectively $\as$-dissolution during heating of Ti-6Al-4V. We thus chose a diffusional nucleation model in Equation \eqref{eqn: alpha_s_diss}
for the resulting $\as \rightarrow \beta$ transformation:
\begin{equation}
    \label{eqn: alpha_s_diss}
    \dXatob =
    \begin{cases}
    k_{\beta}(T) \cdot \left(\tilde{\mathrm{X}}_{\beta} \right)^{\frac{c_{\beta}-1}{c_{\beta}}} \cdot \left(\Xalpha-\Xaeq\right)^{\frac{c_{\beta}+1}{c_{\beta}}} , & \text{for }\ \Xalpha > \Xaeq\\
    0, & \text{else.}
    \end{cases}
\end{equation}
Here, $\tilde{\mathrm{X}}_{\beta}\!=\!\Xb-0.1$ defines a corrected $\beta$-phase fraction. This ensures that the second factor in~\eqref{eqn: alpha_s_diss}, which can be interpreted as a measure for the increasing diffusion interface during the formation process, starts at a value of zero when heating material with initial equilibrium composition $\Xb\!=\!\Xbeq$ above $\Tbs$. By this means, the temporal evolution of $\tilde{\mathrm{X}}_{\beta}$ (i.e., of the additional $\beta$-material beyond $10\%$) begins with a horizontal tangent when exceeding $\Tbs$ as also observable in corresponding experiments \cite{Elmer.2004}. From a physical point of view this experimental observation - as well as the corresponding diffusion model in~\eqref{eqn: alpha_s_diss} - rather correspond to a phase nucleation and subsequent growth process of a new phase fraction $\tilde{\mathrm{X}}_{\beta}$ (starting at $\tilde{\mathrm{X}}_{\beta}=0$) than a continued growth process of a pre-existing phase fraction $\Xb$ (starting at $\Xb\!=\!0.1$). While existing experimental data in terms of temporal phase fraction evolutions is very limited, there is still significant experimental evidence that the $\as \rightarrow \beta$ dissolution should be considered as a (time-delayed) diffusion-based process~\cite{Kelly.Diss, Elmer.2004} rather than an instantaneous transformation, which for simplicity has been assumed in some existing modeling approaches, like e.g., \cite{Murgau.Diss}.\\

Let us again consider the case of lacking $\beta$-phase (i.e. $\Xb\!<\!\Xbeq$), which can only occur in the temperature interval $T \in [\Tbs; \Tbe]$ as discussed in the paragraph above. If in this scenario, a certain amount of remaining Martensite material (i.e. $\Xm>0$) exists, it is assumed that the $\am$-phase fraction is decreased in an instantaneous $\am \rightarrow \beta$ transformation such that $\Xb$ can follow the energetically favorable equilibrium composition $\Xbeq$. Mathematically, this modeling assumption can be expressed by an inequality constraint with the following KKT conditions:
\begin{align}
\label{eqn: alpham_dissolve}
\text{If} \quad \Xm>0: \quad \Xb -\Xbeq \geq 0 \quad \wedge \quad  \dXmtob \geq 0 \quad \wedge \quad (\Xb-\Xbeq) \cdot \dXmtob = 0.
\end{align} 
The constraint $\Xb \!-\! \Xbeq \!\geq\! 0$ (first inequality in~\eqref{eqn: alpham_dissolve}) states that $\Xb$ cannot fall below the equilibrium composition $\Xbeq$ as long as a remainder of Martensite, instantaneously transformable into $\Xb$, is present. Again, the formation rate $\dXmtob$ (second inequality in~\eqref{eqn: alpham_dissolve}) plays the role of a Lagrange multiplier enforcing the constraint $\Xb\!-\!\Xbeq \!=\! 0$ as long as the $\am \rightarrow \beta$ transformation takes place. According to the complementary condition (third equation in~\eqref{eqn: alpham_dissolve}), this formation rate vanishes in case of excessive $\beta$-phase (i.e. $\dXmtob\!=\!0$ if $\Xb\!-\!\Xbeq \!>\! 0$). Since in the considered scenario, the remaining contributions to $\dXb$ vanish, i.e., $\dXatob\!=0\!$ and $\dXbtoa\!=\!0$ due to $\Xb\!=\!\Xbeq$ as well as $\dXbtoam\!=\!0$ due to  $T \!>\! \Tms$,~\eqref{eqn: definition_rates_beta} together with the differential form of the constraint $\dot{\mathrm{X}}_{\beta} \!=\! \dot{\mathrm{X}}_{\beta}^{\text{eq}}$ allows to explicitly determine the corresponding Lagrange multiplier to $\dXmtob\!=\!\dot{\mathrm{X}}_{\beta}^{\text{eq}}$. Once all Martensite is dissolved, i.e., $\Xm\!=\!0$, the $\beta$-phase fraction cannot follow the corresponding equilibrium composition $\Xbeq$ in an instantaneous manner anymore, but rather in a time-delayed manner based on the diffusion process~\eqref{eqn: alpha_s_diss}.\\

To close the system of model equations proposed in this section, a specific expression for the temperature-dependence of the diffusion rates $k_{\as}(T)$ and $k_{\beta}(T)$ required in~\eqref{eqn: alpha_formation_der_new},~\eqref{eqn: alpha_formation_der2_new} and~\eqref{eqn: alpha_s_diss}  has to be made. The mobility of the diffusing species, which is represented by these diffusion rates, is typically assumed to increase with temperature. However, it is assumed that the diffusion rates do not increase in a boundless manner but rather show a saturation at a high temperature level.  Moreover, for the considered class of thermally-activated processes, these diffusion rates are assumed to drop to zero at room temperature. The following type of \emph{logistic functions} represent a mathematical tool for the described system behavior:
\begin{align}
    \label{eqn: diff_rate_model}
    k_{\as}(T) & :=\frac{k_1}{1+\exp\left[-k_3 \cdot \left(T-k_2\right)\right]}
\end{align}
The free parameters $k_1, k_2, k_3$ and $c_{\as}$ governing the $\beta \rightarrow \as$-diffusion processes~\eqref{eqn: alpha_formation_der_new} and~\eqref{eqn: alpha_formation_der2_new} will be inversely determined in Section~\ref{sec: calibration} based on numerical simulations and experimental data for time-temperature-transformations (TTT). The same temperature-dependent characteristic as in~\eqref{eqn: diff_rate_model} is also assumed for the $\as \rightarrow \beta$-diffusion process~\eqref{eqn: alpha_s_diss}. Since the dissolution of $\as$- into $\beta$-phase is reported to take place at higher rates~\cite{Elmer.2004, Kelly.Diss} as compared to the $\as$-formation out of $\beta$-phase, we allow for an increased diffusion rate $k_{\beta}(T)$ of the form:
\begin{align}
    \label{eqn: diff_rate_model2}
    k_{\beta}(T) := f  \cdot k_{\as}(T) \quad \text{with} \quad f>0.
\end{align}
Thus, only the two free parameters $f$ and $c_{\beta}$ are required for the $\as \rightarrow \beta$-diffusion. These two parameters will be inversely determined in Section~\ref{sec: calibration} based on heating experiments taken from~\cite{Elmer.2004}.
 
\begin{remark}[Martensite cooling rate] 
In our model the critical cooling rate $\dTm=-410\ K/s$ is not prescribed as an explicit condition for Martensite formation as done in existing microstructure modeling approaches \cite{Murgau.Diss, Irwin.2017}. Instead, the process of Martensite formation is a pure consequence of physically motivated energy balances and driving forces for diffusion processes. In Section~\ref{sec:cct_validation} it will be demonstrated that a value very close to $\dTm=-410\ K/s$ results from the present modeling approach in a very natural manner when identifying the critical rate for pure Martensite formation from continuous-cooling-transformation (CCT) diagrams created numerically by means of this model.
\end{remark}

\begin{remark}[Johnson-Mehl-Avrami-Kolmogorov (JMAK) equations]
In other publications \cite{Kelly.Diss,Murgau.Diss,Murgau.D,Irwin.2017,Elmer.2004, rae2019thermo,Crespo.2011,lindgren2016simulation}, Johnson-Mehl-Avrami-Kolmogorov (JMAK) equations are used to predict the temporal evolution of the considered phase fractions. It has to be noted that JMAK equations are nothing else than analytic solutions of differential equations (for diffusion processes) very similar to~\eqref{eqn: alpha_formation_der_new}, which are, however, only valid in case of constant parameters $k_{\as}$, $\Xa$ and $\Xbeq$. Since these parameters (due to their temperature-dependence) are not constant for the considered class of melting problems, only a direct solution of the differential equations via numerical integration, as performed in this work, can be considered as mathematically consistent. We furthermore want to note that the mathematical form of JMAK equations, which involve logarithmic and exponential expressions, is prone to numerical instabilities in practical scenarios, especially for the extremely high temperature rates that appear in SLM processes. In contrast, the proposed algorithm in the following Section \ref{sec: microstructure_model_numerical} has a simple and robust mathematical character.\\
\end{remark}

\subsubsection{Temporal discretization and numerical algorithm}
\label{sec: microstructure_model_numerical}
For the numerical solution of the microstructural evolution laws from the previous section, we assume that a temporally discretized temperature field ($T^{n},\dT^{n}$) based on a time step size $\Delta t$ is available at each discrete time step $n$ (e.g. provided by a thermal finite element model as presented in Section~\ref{sec: effective_medium}). In principle, any time integration scheme can be employed for temporal discretization of the phase fraction evolution equations from the last section. Specifically, in the subsequent numerical examples either an implicit Crank-Nicolson scheme or an explicit forward Euler scheme have been applied. For simplicity, the general algorithmic realization of the time-discrete microstructure evolution model is demonstrated on the basis of a forward Euler scheme.

For the following time integration procedure of~\eqref{eqn: definition_rates} it has to be noted that only the rates $\dot{\mathrm{X}}_{\am \rightarrow \as},\dot{\mathrm{X}}_{\beta \rightarrow \as},\dot{\mathrm{X}}_{\as\rightarrow \beta}$ corresponding to diffusion processes will be integrated in time. Instead of integrating the rates $\dXbtoam$ and $\dXmtob$ corresponding to instantaneous Martensite formation and dissolution processes, the associated constraints in~\eqref{eqn: martensite-formation} and~\eqref{eqn: alpham_dissolve} will be considered directly by means of algebraic constraint equations. In a first step, assume that the temperature data $T^{n+1},\dT^{n+1}$ of the current time step $n+1$ as well as the microstructure data $\Xa^{n},\Xm^{n},\dot{\mathrm{X}}_{\am \rightarrow \as}^{n},\dot{\mathrm{X}}_{\beta \rightarrow \as}^{n},\dot{\mathrm{X}}_{\as\rightarrow \beta}^{n}$ of the last time step $n$ is known. With this data, the microstructure update is performed:
\begin{align}
    \label{eqn: alphas_update}
    \Xa^{n+1} \!=\! \Xa^{n} + \Delta t (\dot{\mathrm{X}}_{\beta \rightarrow \as}^{n}+\dot{\mathrm{X}}_{\am \rightarrow \as}^{n} - \dot{\mathrm{X}}_{\as\rightarrow \beta}^{n}) \quad \text{and} \quad \Xm^{n+1} \!=\! \Xm^{n} - \Delta t \dot{X}_{\am \rightarrow \as}^{n}.
\end{align}
The time integration error in~\eqref{eqn: alphas_update} might lead to a violation of the scope $\Xa^{n+1}, \Xm^{n+1} \in [0;0.9]$ of the phase fraction variables. In this case the relevant phase fraction variable is simply limited to its corresponding minimal or maximal value, respectively. Similarly, if $\Xalpha^{n+1}=\Xa^{n+1}+\Xm^{n+1}$ exceeds the maximum value of $0.9$ the individual contributions $\Xa^{n+1}$ and $\Xm^{n+1}$ are reduced such that the maximum value $\Xalpha^{n+1}=0.9$ is met and the ratio $\Xa^{n+1}/\Xm^{n+1}$ is preserved. Subsequently, the $\beta$-phase fraction is calculated from the continuity equation $\Xb^{n+1}=1-\Xalpha^{n+1}$. Afterwards, the updated equilibrium phase fractions $\Xa^{\text{eq},n+1}$ and $\Xm^{\text{eq},n+1}$ are calculated according to~\eqref{eqn: mean_approx}-\eqref{eqn: am_eq_general} with $\Xa^{n+1}$ and $T^{n+1}$ before the corresponding equilibrium composition $\Xb^{\text{eq},n+1}=1-\Xa^{\text{eq},n+1}$ of the $\beta$-phase is updated. Next, a potential instantaneous Martensite formation out of the $\beta$-phase according to~\eqref{eqn: martensite-formation} is considered as follows:
\begin{subequations}
\begin{align}
    \label{eqn: alpham_formation}
   \text{If } \Xm^{n+1} < {\Xmeq}^{,n+1}, \quad \text{Update: }& \Xb^{n+1} \leftarrow \Xb^{n+1} +  \Xm^{n+1} -  {\Xmeq}^{,n+1},\\
   																				     \text{Set: }& \Xm^{n+1}={\Xmeq}^{,n+1}.
\end{align}
\end{subequations}
Similarly, a potential instantaneous Martensite dissolution into $\beta$-phase according to~\eqref{eqn: alpham_dissolve} is considered as follows:
\begin{subequations}
\begin{align} 
    \label{eqn: alpham_formation}
   \text{If } \Xb^{n+1} < {\Xbeq}^{,n+1} \wedge \Xm^{n+1}>0, \quad \text{Update: }& \Xm^{n+1} \leftarrow \Xm^{n+1} +  {\Xbeq}^{,n+1} -  \Xb^{n+1},\\
   																									                  \text{Set: }& \Xb^{n+1}={\Xbeq}^{,n+1}.
\end{align}
\end{subequations}
Again, if necessary the increment in~\eqref{eqn: alpham_formation} is limited such that the updated phase fraction $\Xm^{n+1}$ does not become negative. As a last step, the diffusion-based transformation rates $\dot{\mathrm{X}}_{\beta \rightarrow \as}^{n+1}$, $\dot{\mathrm{X}}_{\am \rightarrow \as}^{n+1}$ and $\dot{\mathrm{X}}_{\as\rightarrow \beta}^{n+1}$ according to~\eqref{eqn: alpha_formation_der_new},~\eqref{eqn: alpha_formation_der2_new} and~\eqref{eqn: alpha_s_diss}, all evaluated at time step $n+1$, are calculated. With these results, the next time step $n+2$ can be calculated starting again with~\eqref{eqn: alphas_update}.\\

 \begin{remark}[Initial conditions for explicit time integration]
 While the diffusion process according to Equation~\eqref{eqn: alpha_formation_der2_new} will start at a configuration with $\Xm \neq 0$ and $\Xa \neq 0$, Equation~\eqref{eqn: alpha_formation_der_new} needs to be evaluated for $\Xa = 0$ to initiate the diffusion process. However, the evolution of Equation~\eqref{eqn: alpha_formation_der_new} based on an explicit time integration scheme will remain identical to zero for all times for a starting value of $\Xa = 0$. Therefore, during the first cooling period the phase fraction has to be initialized at the first time step $t^n$ where $\Xaeq>0$. In the following, the initialization procedure considered in this work is briefly presented. In a first step,~\eqref{eqn: alpha_formation_der_new} is reformulated using the relations $\Xb=1-\Xalpha$ and $\Xbeq=1-\Xaeq$ as well as the approximate assumptions $k_{\as}(T)=const.$, $\Xaeq=const.$ and $\Xm=0$ (i.e., $\Xalpha=\Xa$) for the initial state \cite{avramov2014generalized}:
  \begin{align}
     \label{eqn: alpha_formation_der_new_nils_reformulated}
     \dot{g}=\tilde{k} \cdot g^{\!\frac{c_{\as}-1}{c_{\as}}\!} \cdot (1-g)^{\!\frac{c_{\as}+1}{c_{\as}}} \quad \text{with} \quad
     g= \frac{\Xalpha}{\Xaeq}, \quad 
     \tilde{k}=k_{\as} \cdot \Xaeq.
 \end{align}
 Based on the analytic solution in \cite{avramov2014generalized}, evaluated after one time step $\Delta t$, the initialization for $\Xa^{n}$ at $t^n$ reads:
 \begin{align}
     \label{eqn: alpha_formation_der_new_nils_analytic_init}
     \Xa^{n}=\Xaeq\cdot
     \left[\,
     1+
     \left(
     \frac{c_{\as}}{\tilde{k} \Delta t}
     \right)^{c_{\as}} \,\, 
     \right]^{-1}.
 \end{align}
 We compared this approach with an implicit Crank-Nicolson time integration (either used for the entire simulation or only for initialization of the first time step), where the initial $\as$-phase fraction $\Xa$ does not need to be set explicitly and found no differences in the resulting diffusion dynamics according to~\eqref{eqn: alpha_formation_der_new}.
 \end{remark}

\section{Inverse parameter identification and validation of microstructure model}
\label{sec: calibration}

The four parameters $\boldsymbol{\theta}_{\text{diff},\as}=[c_{\as}, k_1, k_2,k_3]^T$ according to Equations~\eqref{eqn: alpha_formation_der_new},~\eqref{eqn: alpha_formation_der2_new} and~\eqref{eqn: diff_rate_model} as well as the two parameters $\boldsymbol{\theta}_{\text{diff},\beta}=[c_{\beta}, f]^T$ according to Equations \eqref{eqn: alpha_s_diss} and \eqref{eqn: diff_rate_model2}, are so far still unknown and need to be inversely identified via experimental data sets. For the inverse identification of $\boldsymbol{\theta}_{\text{diff},\as}$, we use so-called time-temperature-transformation (TTT) experiments, a well-known experimental characterization procedure for microstructural evolutions \cite{Murgau.Diss, rae2019thermo, Kelly.Diss, enns1983time}. As TTT-experiments only capture the dynamics of cooling processes, we will identify the parameters $\boldsymbol{\theta}_{\text{diff},\beta}$ governing the heating dynamics of the microstructure via data from heating experiments taken from~\cite{Elmer.2004}.

\subsection{Inverse identification of $\as$-formation dynamics via TTT-data}
Time-temperature transformation (TTT) experiments \cite{bhadeshia1982thermodynamic} are one of the most important and established procedures for (crystallographic) material characterization. The goal of the TTT investigations is to understand the \emph{isothermal} transformation dynamics of an alloy by plotting the percentage volume transformation of its crystal phases over time. Thereto, the material is first equilibriated at high temperatures such that only the high-temperature phase is present. Afterwards, the material is rapidly cooled down to a target temperature at which it is then held constant over time so that the isothermal phase transformation at this temperature can be recorded. Rapid cooling refers here to a cooling rate that is fast enough so that diffusion-based transformations during the cooling itself can be neglected and can subsequently be studied under isothermal conditions at the chosen target temperature. The procedure is repeated for successively reduced target temperatures. The emerging diagram of phase contour-lines over the $T\times\log(t)$ space is commonly referred to as TTT-diagram.

\begin{figure}[htbp]
        \centering
        \includegraphics[scale=0.35]{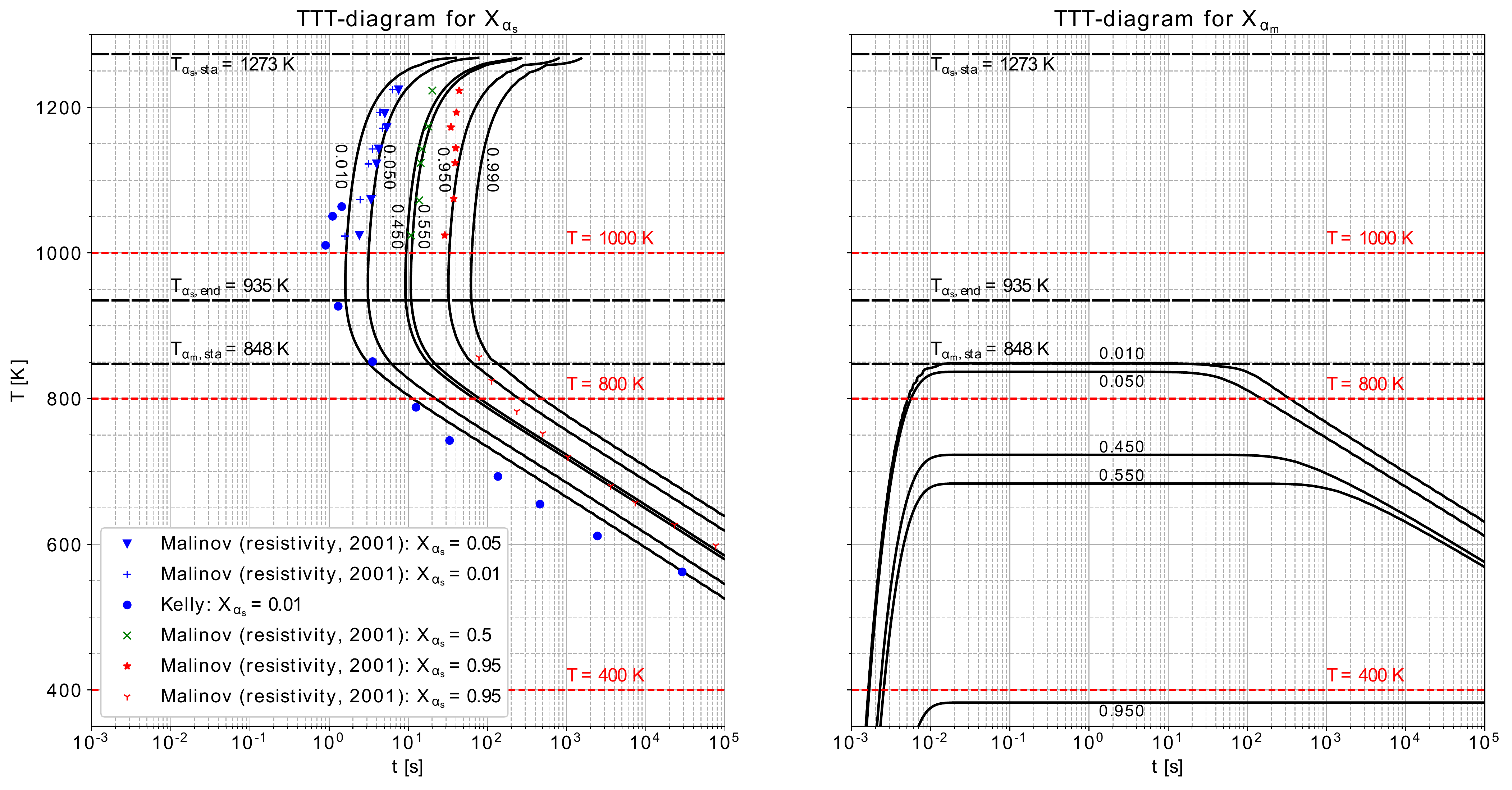}
         \caption{Simulation of the TTT-diagram for the $\as$- and $\am$-phases using the maximum likelihood point estimate for the uncertain kinetic parameters $\boldsymbol{\theta}_{\text{diff},\as}^*$ of the microstructure evolution, along with experimental data by Malinov \cite{Malinov.2001.Resistivity} and Kelly \cite{Kelly.article}. Left: Contour-lines for $\Xa$; Right: Contour-lines for $\Xm$. Contour lines are shown for the $1\%$, $5\%$, $45\%$, $55\%$, $95\%$ and $99\%$ normalized phase fractions. Three temperatures are marked in red and discussed in the analysis.} 
        \label{fig: TTT_diagram}
    \end{figure}

In the present work, the simulation of TTT-curves for Ti-6Al-4V was conducted as follows: We initialized the microstructure state at $T=1400\ K > \Tbs$ with pure $\beta$-phase, such that $\Xb=1.0$. Afterwards, the microstructure was quickly cooled down with $\dT=-500\ K/s$\footnote{We chose a constant cooling rate here as the detailed cooling dynamics are not important for TTT diagrams as long as cooling takes place "fast enough". In the latter respect, we also investigated several higher cooling rates without observing differences in the resulting TTT-diagrams.} to a target temperature $T_{\text{target}}\in[350,\ 1300]$ and the evolution of the microstructure at this target temperature was recorded over time. The range of target temperatures was discretized in steps of $10 \ K$ such that 95 individual target temperatures and hence microstructure simulations were considered. Figure \ref{fig: TTT_diagram} depicts the isolines of simulated $\Xa$- (left) and $\Xm$-phase fractions (right), after identifying the model parameters via the experimental data \cite{Murgau.Diss, rae2019thermo, Kelly.Diss, Malinov.2001.Resistivity} shown in the left figure.
Please note that the phase-fractions in Figure \ref{fig: TTT_diagram} were normalized with $\Xaeq(T)$, which takes on a value of $0.9$ for temperatures below $\Tbs$, as this was also the case in the underlying experimental investigations. We highlighted three temperatures $T=\{400,800,1000\}\ K$ to discuss the microstucture evolution at these points. 

\begin{figure}[htbp]
    \centering
    \includegraphics[scale=0.3]{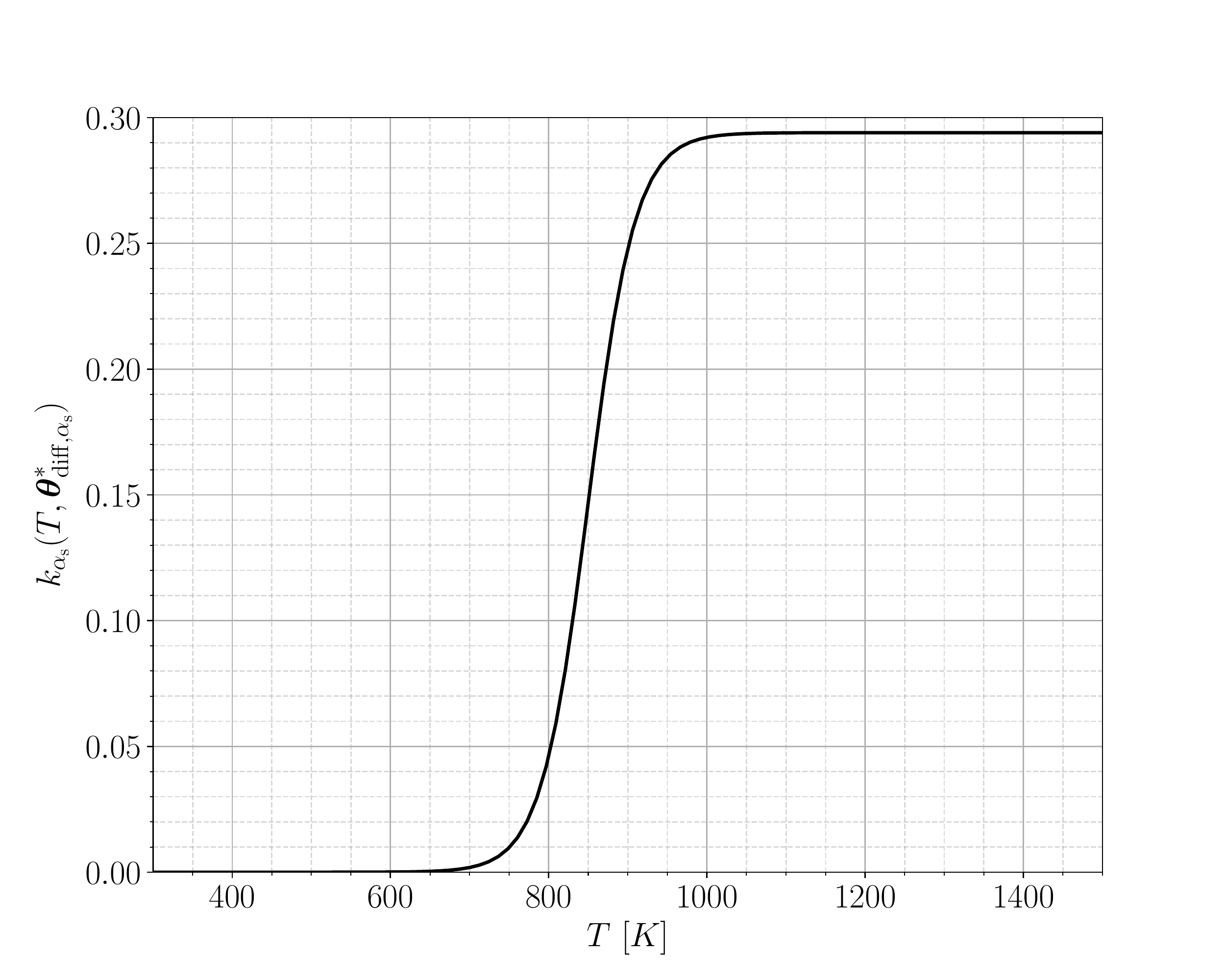}
     \caption{Temperature-dependent diffusion rate $k_{\as}(T, \boldsymbol{\theta}_{\text{diff},\as}^*)$ for $\as$-formation resulting from inverse parameter identification.} 
    \label{fig: k_rate_T}
\end{figure}

At $T=1000\ K$ (between $\Tbe$ and $\Tbs$, i.e. above $\Tms$) we get a pure $\beta\rightarrow\as$ transformation. When looking at higher target temperatures $T>1000\ K$ it can be observed that the isoline with 1\% phase fraction for $\Xa$ is shifted to longer times, which results from a decreasing value of $\Xaeq$ (i.e. a decreasing driving force) that slows down the initial dynamics of $\Xa$-formation at elevated temperatures, even-though $k_{\as}$ is already saturated at its maximal value in this temperature range (see Figure~\ref{fig: k_rate_T}). At $T=800\ K$ we are now below $\Tms$, so that the initial cool-down results (instantaneously) in a Martensite phase fraction according to $\Xmeq$. With ongoing waiting time, the remaining $\beta$-phase transforms in a diffusion-driven manner into stable $\alpha_s$-phase. In the TTT-diagram, we can already notice that the isoline with 1\% phase fraction for $\Xa$ is shifted to longer times as compared to the higher temperature level $T=1000\ K$, which, this time, is caused by the decreasing value of $k_{\as}$ for lower temperatures, as depicted in Figure \ref{fig: k_rate_T} and modeled in Equation \eqref{eqn: diff_rate_model}. Moreover, the right-hand side of Figure \ref{fig: TTT_diagram} shows that the Martensite phase fraction decreases again for waiting times $t>100s$, which represents the diffusion-based dissolution of Martensite into $\alpha_s$-phase according to~\eqref{eqn: alpha_formation_der2_new}.

Finally, the transformation at $T=400\ K$ initially results in almost the maximal possible amount of Martensite (Remember: The $100\%$-isoline in Figure \ref{fig: TTT_diagram} corresponds to a phase fraction of $0.9$ for $T<\Tbs$). As the diffusion rate $k_{\as}$ is almost zero for such low temperatures, the Martensite phase cannot be dissolved to stable $\as$ in finite times and the Martensite phase remains present as metastable phase at low temperatures, which agrees well with experimental observations. In the TTT-diagram this effect shifts the isolines asymptotically to $t\rightarrow \infty$ when approaching the room temperature. All in all, the shift to longer times due to a low value of $k_{\as}$ for low temperatures and the delay effect due to a decreasing (driving force) value of $\Xaeq$ at high temperatures leads to the typical C-shape of TTT-phase-isolines.

Inverse parameter identification was conducted for the diffusion parameters $\boldsymbol{\theta}_{\text{diff},\as}$ by maximizing the data's likelihood for $\boldsymbol{\theta}_{\text{diff},\as}$. We assumed a (conditionally independent) static Gaussian noise for the measurements on the $\log(t)$-scale. This assumption is equivalent to a log-normal distributed noise in the data along the time-scale. The maximum-likelihood point estimate can then be determined by solving the following least-square optimization problem (see Remark below for more details):      
\begin{align}
    \label{eqn: inverse_TTT}
    \begin{split}
    \boldsymbol{\theta}_{\text{diff},\as}^{*} &= \underset{\boldsymbol{\theta}_{\text{diff},\as}}{\mathrm{argmin}}\sum\limits_{i} \bigg({\Xa}_{\mathrm{,TTT}}\left(\log(t_{i}),T_{i}, \boldsymbol{\theta}_{\text{diff},\as}\right)-{\Xa}_{,\mathrm{TTT, exp.},i}\bigg)^2\\
    \end{split}
\end{align}
In Equation \eqref{eqn: inverse_TTT} the term ${\Xa}_{\mathrm{,TTT}}\left(\log(t_i),T_i, \boldsymbol{\theta}_{\text{diff},\as}\right)$ describes the simulated TTT-phase fractions (normalized by $\Xaeq(T)$) at time $t_i$, Temperature $T_i$ and for diffusional parameters $\boldsymbol{\theta}_{\text{diff},\as}$. The index $i$ marks here the specific temperatures and times for which the corresponding observed experimental data ${\Xa}_{,\mathrm{TTT, exp.},i}$ was recorded.

We utilized a Levenberg-Marquardt optimization routine \cite{more1978levenberg}, which is implemented in our in-house software framework \textit{QUEENS} \cite{queens} to iteratively solve Equation \eqref{eqn: inverse_TTT}. The result of this optimization procedure is given by the parameter set ${\boldsymbol{\theta}_{\text{diff},\as}^*=[c_{\as},k_{1},k_2,k_3]^T=[2.51,\ 0.294,\ 850.0,\ 0.0337]^T}$.
Additionally, Figure \ref{fig: k_rate_T} visualizes the temperature-dependent diffusion rate $k_{\as}(T, \boldsymbol{\theta}_{\text{diff},\as}^*)$ according to Equation \eqref{eqn: diff_rate_model} that results from these parameters.

\FloatBarrier

\begin{remark}[Assumptions for inverse identification]
The least-squares optimization problem in Equation \eqref{eqn: inverse_TTT} is the result of the following assumptions for the inverse identification task of $\boldsymbol{\theta}_{\text{diff},\as}^{*}$:
The simulation model is expressed by a mapping $y(\boldsymbol{\theta},\boldsymbol{\mathfrak{c}})$, with $\boldsymbol{\theta}$ being model parameters that should be inferred from data via inverse analysis and $\boldsymbol{\mathfrak{c}}$ being coordinates on which the model output is recorded. Further model inputs (e.g., parameters that we want to keep fixed) are omitted to avoid cluttered notation.
We assume that a vector of $n_{\text{obs}}$ scalar experimental observations $\boldsymbol{y}_{\text{obs},\boldsymbol{\mathfrak{C}}}$, recorded at coordinates $\boldsymbol{\mathfrak{C}}=\{\boldsymbol{\mathfrak{c}}_i\}$, was disturbed by conditionally independent and static Gaussian noise with variance $\sigma_n^2$ on the $\log(t)$-space. This assumption is equivalent to log-normal distributed noise on the $t$-space.
Statistically, the former assumption is expressed by the so called \emph{likelihood function} $l(\boldsymbol{\theta})=p(\boldsymbol{y}_{\text{obs},\boldsymbol{\mathfrak{C}}}|y(\boldsymbol{\theta},\boldsymbol{\mathfrak{C}}))$, which is the probability density for the observed data $\boldsymbol{y}_{\text{obs},\boldsymbol{\mathfrak{C}}}$, given a specific choice of the parameterized simulation model $y(\boldsymbol{\theta},\boldsymbol{\mathfrak{C}})$, evaluated at the same coordinates $\boldsymbol{\mathfrak{C}}$. Gaussian conditional independent noise implies now that (theoretically) repeated observations of experimental data $\boldsymbol{y}_{\text{obs},\boldsymbol{\mathfrak{c}_i}}$ at each location $\boldsymbol{\mathfrak{c}}_i$ will result in $n_{\text{obs}}$ Gaussian probability distributions with variance $\sigma_n^2$ for $y_{\text{obs},\boldsymbol{\mathfrak{c}}_i}$. Independence in particular means that a disturbance by noise at $\boldsymbol{\mathfrak{c}_i}$ does not influence the disturbance by noise at any other point. Ideally, the simulation output should reflect the mean of the observation noise at each $\boldsymbol{\mathfrak{c}_i}$. The latter can be expressed by the product (due to the conditional independence) of $n_{\text{obs}}$ Gaussian distributions with their mean values being the simulation outputs $y(\boldsymbol{\theta},\boldsymbol{\mathfrak{c}}_i)$ at each $\boldsymbol{\mathfrak{c}}_i$ and  with variance $\sigma_n^2$:
$l(\boldsymbol{\theta})=p(\boldsymbol{y}_{\text{obs},\boldsymbol{\mathfrak{C}}}|y(\boldsymbol{\theta},\boldsymbol{\mathfrak{C}}))=\prod\limits_{i=1}^{n_{\text{obs}}} \mathcal{N}\left(y_{\text{obs},\mathfrak{c}_i}|y(\boldsymbol{\theta},\mathfrak{c}_i),\sigma_n^2\right)\propto \exp\left[-\frac{\sum_i(y_{\text{obs},\mathfrak{c}_i}-y(\boldsymbol{\theta},\mathfrak{c}_i))^2}{\sigma^2_n}\right]$. Despite $l(\boldsymbol{\theta})$ being a true probability density function for $\boldsymbol{y}_{\text{obs},\boldsymbol{\mathfrak{C}}}$, the latter is not the case w.r.t. the model parameters $\boldsymbol{\theta}$, such that $l(\boldsymbol{\theta})$ is mostly referred to as the \emph{likelihood function}. The maximum likelihood (ML) point estimate $\boldsymbol{\theta}^{*}$ refers then to a value of $\boldsymbol{\theta}$ that maximizes the likelihood function $l(\boldsymbol{\theta})$, respectively the probability density of the observations $\boldsymbol{y}_{\text{obs},\boldsymbol{\mathfrak{C}}}$ for the specific choice $\boldsymbol{\theta}^{*}$. The maximum of $l(\boldsymbol{\theta})$ can be found by minimizing the sum of the square terms in the argument of the exponential function, which leads ultimately to Equation \ref{eqn: inverse_TTT}. 
\end{remark}

\subsection{Inverse identification and validation of diffusional heating dynamics}
\begin{figure}[htbp]
        \centering
        \includegraphics[scale=0.27]{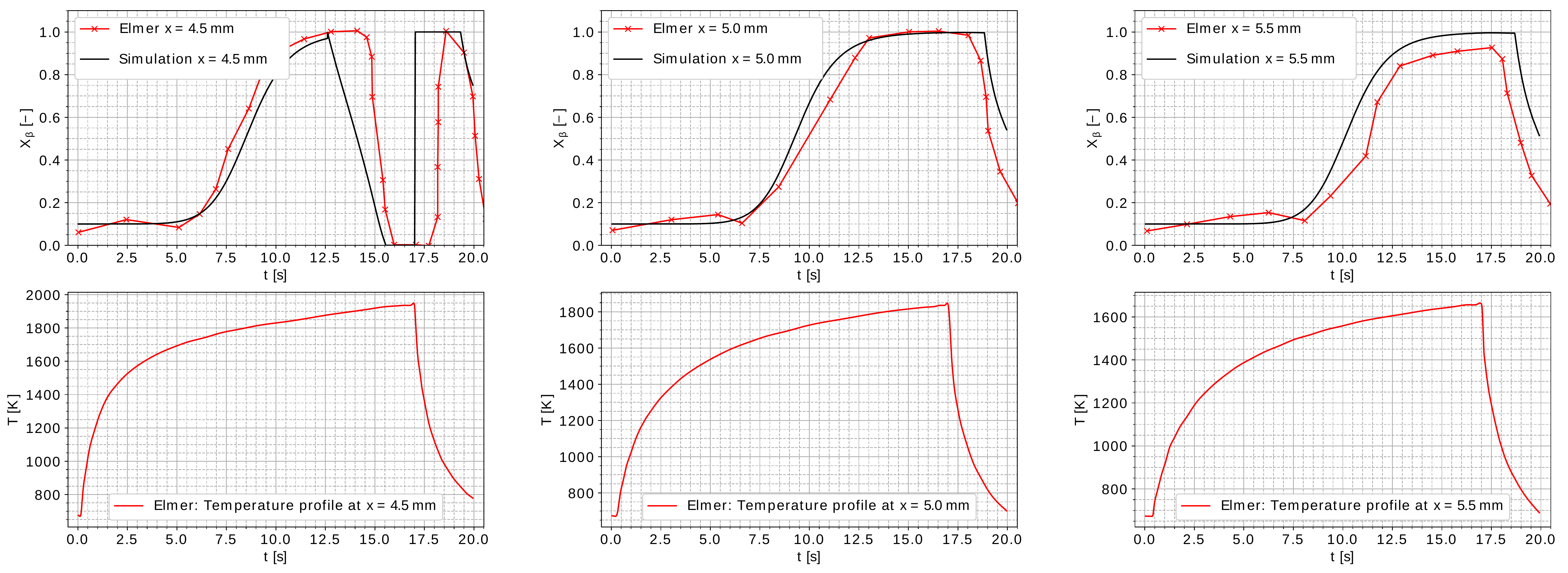}
         \caption{Experimental measurements for $\beta$-phase evolution in a welding process of Ti-6Al-4V by Elmer et al. \cite{Elmer.2004} along with the corresponding simulation results based on the proposed microstructure model and the identified parameter set $\boldsymbol{\theta}_{\text{diff},\beta}=\boldsymbol{\theta}_{\text{diff},\beta}^*$.} 
        \label{fig: Elmer_fit}
    \end{figure}

So far, the inverse identification of $\boldsymbol{\theta}_{\text{diff},\as}$ via TTT-data only accounts for the cooling dynamics of the microstructure model.  In the following, we inversely identify the parameter set $\boldsymbol{\theta}_{\text{diff},\beta}$ via experimental data from \cite{Elmer.2004}, which consists of three temperature and $\beta$-phase data-sets for three positional measurements for an electron beam welding process with Ti-6Al-4V. Note that the specific measurement positions $x=\{4.5,5.0,5.5\}\ mm$ defined in the original work \cite{Elmer.2004} are not relevant here since we only aim at correlating temperature and phase fraction data. Please note also that this identification step reuses $\boldsymbol{\theta}_{\text{diff},\as}^*$, respectively the temperature characteristics of $k_{\as}$ as identified above and only scales the latter by the factor $f$ (see Equation \eqref{eqn: diff_rate_model2}). The maximum likelihood estimate $\boldsymbol{\theta}_{\text{diff},\beta}^{*}$ was again calculated by solving a least-squares optimization problem with the Levenberg-Marquardt optimizer:
\begin{align}
    \label{eqn: cost_fun_heating}
    \begin{split}
       \boldsymbol{\theta}_{\text{diff},\beta}^{*} = \underset{\boldsymbol{\theta}_{\text{diff},\beta}}{\mathrm{argmin}}\bigg[ &\int\left(\Xb\left(t,T_{\text{exp.},x=4.5}(t),\boldsymbol{\theta}_{\text{diff},\beta}\right)-{\Xb}_{,\mathrm{exp.},x=4.5}(t)\right)^2 dt\\
       + &\int\left(\Xb\left(t,T_{\text{exp.},x=5.0}(t),\boldsymbol{\theta}_{\text{diff},\beta}\right)-{\Xb}_{,\mathrm{exp.},x=5.0}(t)\right)^2 dt\\
        + &\int\left(\Xb\left(t,T_{\text{exp.},x=5.5}(t),\boldsymbol{\theta}_{\text{diff},\beta}\right)-{\Xb}_{,\mathrm{exp.},x=5.5}(t)\right)^2 dt \bigg]
       \end{split}
\end{align}
Equation \eqref{eqn: cost_fun_heating} accounts for all three phase-temperature measurements at locations $=\{4.5,5.0,5.5\}\ mm$ simultaneously leading to a robust point estimate $\boldsymbol{\theta}_{\text{diff},\beta}^{*}$ with low generalization error. The minuends $\Xb\left(t,T_{\text{exp.},x=4.5}(t),\boldsymbol{\theta}_{\text{diff},\beta}\right),\Xb\left(t,T_{\text{exp.},x=5.0}(t),\boldsymbol{\theta}_{\text{diff},\beta}\right)$ and $\Xb\left(t,T_{\text{exp.},x=5.5}(t),\boldsymbol{\theta}_{\text{diff},\beta}\right)$ reflect the simulations results for the $\beta$-phase fraction over time, for the experimental temperature profiles $T_{\text{exp.},x=4.5}(t),T_{\text{exp.},x=5.0}(t)$ and $T_{\text{exp.},x=5.5}(t)$, respectively (see Figure \ref{fig: Elmer_fit} second row). The subtrahends ${\Xb}_{,\mathrm{exp.},x=4.5}(t),{\Xb}_{,\mathrm{exp.},x=4.5}(t)$ and ${\Xb}_{,\mathrm{exp.},x=5.5}(t)$ are the corresponding experimentally measured $\beta$-phase fraction profiles over time.

The inverse identification yields the parameter set $\boldsymbol{\theta}_{\text{diff},\beta}^{*}=[c_\beta,f]^T=[11.0, 3.8]^T$. The comparison of experimental data from \cite{Elmer.2004} and the simulation-based prediction in Figure \ref{fig: Elmer_fit} show very good agreement.
We summarize that the dissolution of the $\as$-phase in the case of $\Xa>\Xaeq$ can be successfully modeled by a diffusional approach for $\as$-nucleation by scaling $k_{\as}$ with a factor of $f^*=3.8$ and selecting $c_\beta^*=11.0$. We will use this approach in the subsequent numerical demonstrations.

\FloatBarrier

\subsection{Validation of calibrated microstructure model via CCT-data}
\label{sec:cct_validation}
\begin{figure}[htbp]
    \centering
    \includegraphics[scale=0.26]{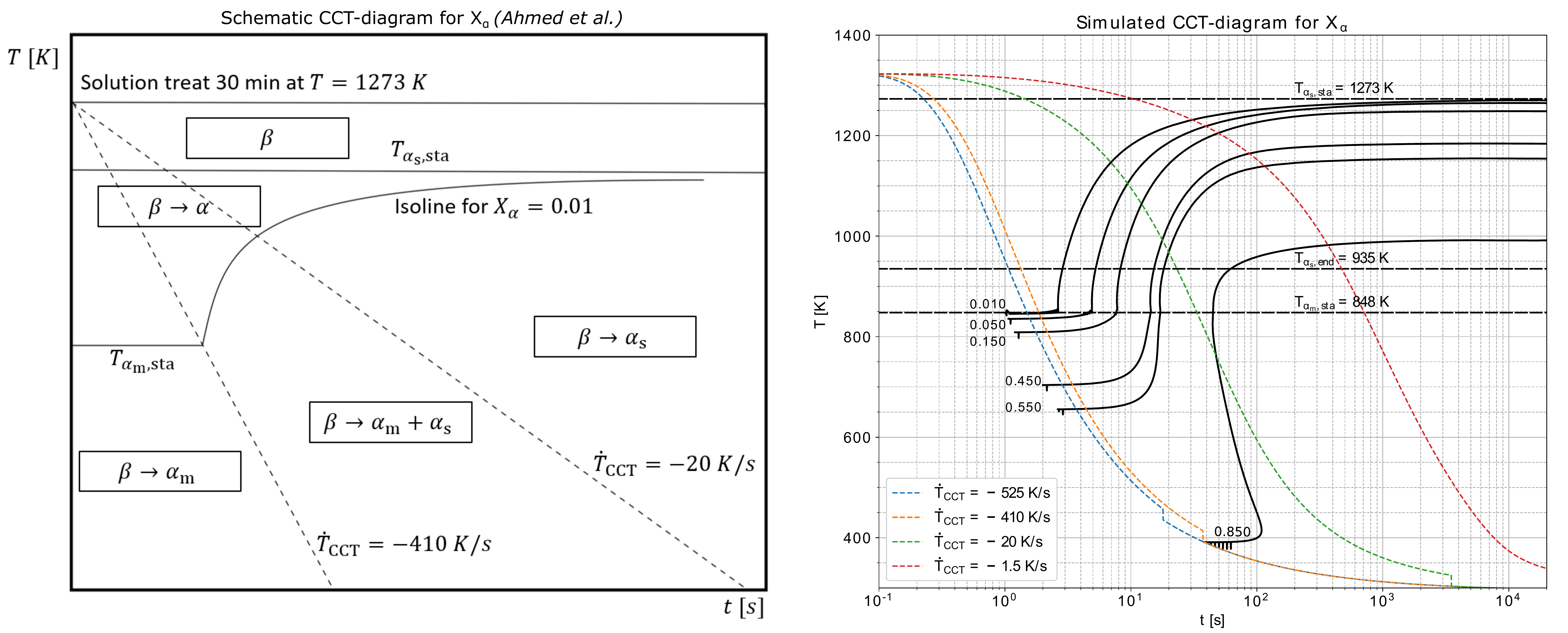}
     \caption{Continuous-cooling transformation (CCT) diagram for $\Xalpha$ in Ti-6Al-4V with characteristic cooling rates $\dot{T}_{\text{CCT}}$: schematic diagram adapted from~\cite{Ahmed.1998} (left) and predicted diagram via proposed microstructure model (right).} 
    \label{fig: CCT_Ahmed}
\end{figure}

To validate the calibrated microstructure model, we will now compare the predicted microstructure evolutions with further experimental data. Another common experimental approach for the characterization of microstructural phase evolutions are the so-called continuous-cooling transformation (CCT) experiments \cite{cahn1956transformation, Ahmed.1998}. Here, the microstructural probe is again equilibriated at a temperature above $\Tbe=1273\ K$, such that $\Xb=1.0$. Afterwards, the probe is cooled down to room temperature $T_{\infty}=293.15\ K$ at different cooling rates $\dot{T}_{\text{CCT}}$. Note that the true cooling rate $\dT(t)$ is time-dependent and $\dot{T}_{\text{CCT}}$ is only a representative descriptor. Eventually, the evolving microstructure of these cooling procedures is recorded on the $t\times T$-space. In contrast to TTT-experiments, the dynamics of the cooling process itself and the thereof resulting microstuctural phase transformations are now the main aspect of these experimental procedures. Hence, the particular temperature profile of the cooling process is now of great importance and has a large impact on the emerging phases.
The true cooling rates $\dT(t)$ in practical CCT-experiments are changing over time, being highest in the beginning of the cooling procedure and tending to zero for infinite long times. Following the procedure in~\cite{Ahmed.1998}, the characteristic cooling rate $\dot{T}_{\text{CCT}}$ is in the following defined as the cooling rate at $900\ ^\circ C$, respectively $1173.15\ K$ $\left(\dot{T}_{\text{CCT}}:=\dot{T}\big |_{T=1173.15\ K}\right)$. A schematic CCT-diagram for Ti-6Al-4V based on characteristic experimental cooling rates taken from~\cite{Ahmed.1998} as well as the prediction by the proposed microstructure model are shown in Figure \ref{fig: CCT_Ahmed}.

In \cite{Ahmed.1998} two distinct CCT-cooling rates $\dot{T}_{\text{CCT}}$ are reported which divide the evolving microstructure for continuous Ti-6Al-4V cooling into three characteristic regimes. For cooling faster than $\dot{T}_{\text{CCT}}=-410 \ K/s$, only martensitic transformation was observed. For cooling between $\dot{T}_{\text{CCT}}=-410 \ K/s$ and $\dot{T}_{\text{CCT}}=-20 \ K/s$, coexisting Martensite and stable $\as$ transformation were found. These coexisting $\alpha$-morphologies are sometimes also described as massive-$\alpha$ \cite{Kelly.Diss}. Eventually, for cooling rates slower than $\dot{T}_{\text{CCT}}=-20 \ K/s$ only stable $\as$ emerges. In the regime of fast cooling rates, the transformation of Martensite starts at temperatures below the Martensite-start-temperature $\Tms$. For slower cooling rates stable $\as$-nucleation is the dominating effect, which can already be observed at higher temperatures. From Figure~\ref{fig: CCT_Ahmed} it can already be concluded that the characteristic kink in the isoline $\Xalpha=0.01$ at $\dot{T}_{\text{CCT}}=-410 \ K/s$, which separates the pure Martensite from the massive-$\alpha$ regime, is predicted very well by the proposed model. A more detailed discussion and comparison will be presented in the following sections.

\FloatBarrier
\subsubsection{Inverse determination of cooling profiles}
In the CCT-simulations presented in the subsequent section, we followed the experimental set-up and results presented in \cite{Ahmed.1998}. In their work, the authors realized different cooling rates through convectional air cooling for low and moderate cooling rates and through water quenching for high cooling rates. The material probe in~\cite{Ahmed.1998} was a Ti-6Al-4V Jominy end quench bar \cite{newkirk2000jominy} of  $31.8\ mm$ diameter cooled down by the mentioned media only at the front face side and thermally isolated at all remaining surfaces. We want to note that analogous CCT-investigations for steel, based on the numerical simulation of Jominy end quench tests, have been conducted in the past \cite{homberg1996numerical}, nevertheless using a simpler microstructure model. To mimic the temperature evolution of the actual experiments, we approximated the temperature profile by the analytic solution for a semi-infinite solid body (sib) under surface convection, which is given in \cite{rohsenow1998handbook}:
\begin{equation}
    \label{eqn: semi_infinite_cooling}
    T_{\text{sib}}(x,t)=(T_{\infty}-T_0) \cdot \left[\text{erfc}\left(\frac{x}{2\sqrt{\alpha\cdot t}}\right) - \exp\left(g\cdot x+g^2\cdot\alpha \cdot t\right)\cdot\text{erfc}\left(\frac{x}{2\sqrt{\alpha t}}+g\cdot\sqrt{\alpha t}\right)\right] + T_0
\end{equation}

\begin{figure}[htbp]
    \centering
    \includegraphics[scale=0.4]{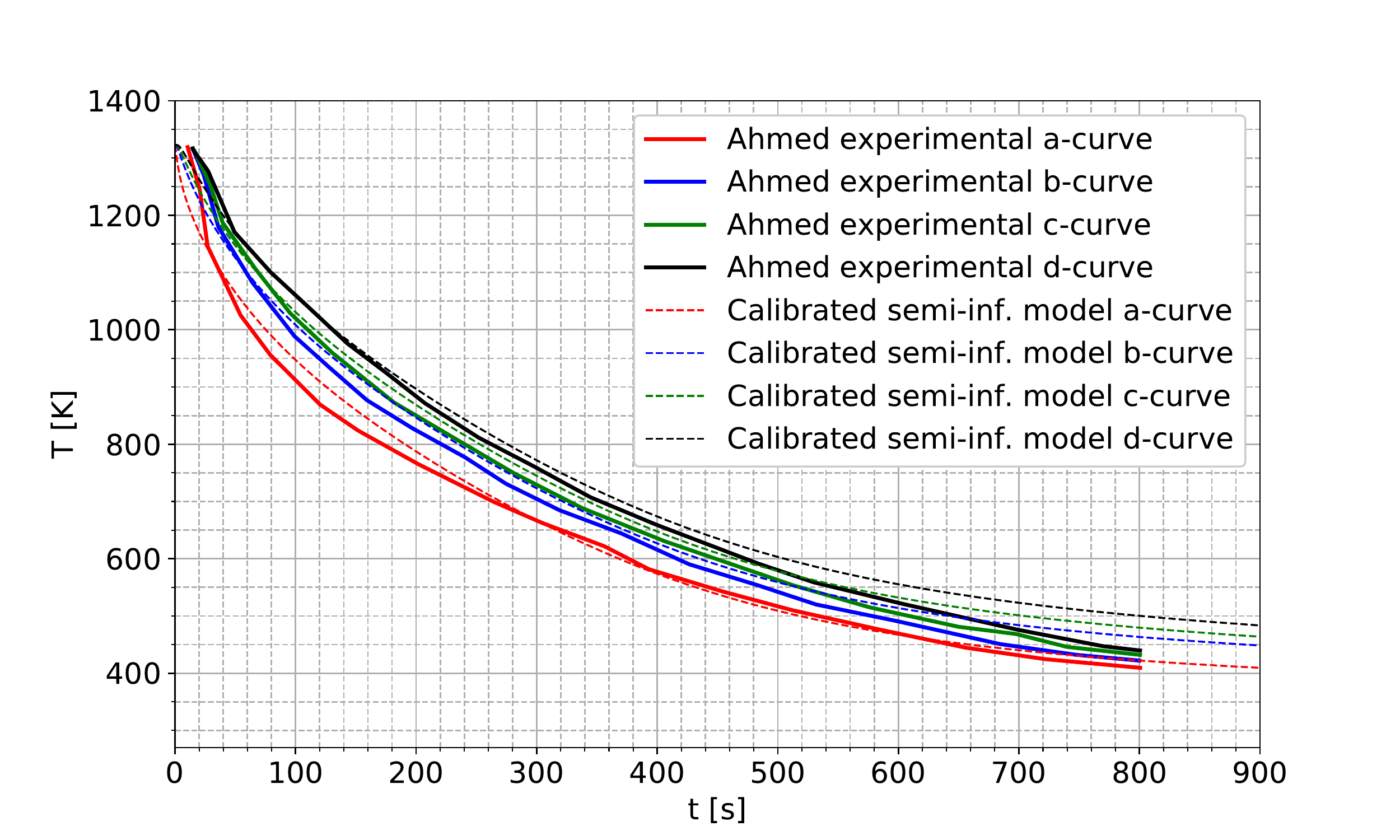}
     \caption{Measured continuous cooling curves by Ahmed et al.  \cite{Ahmed.1998} along with the corresponding model-based curves resulting from the identified temperature-dependent parameter $g(T)$ from Equations \eqref{eqn: quadratic_heuristic} and \eqref{eqn: semi_infinite_cooling}. The a-, b-, c- and d-curve were measured at positions $x_a=3.2\ mm$, $x_b=9.5\ mm$, $x_c=12 \ mm$ and $x_d=15.2\ mm$ at otherwise identical cooling conditions.} 
    \label{fig: semi_inf_cooling}
\end{figure}

Here, $\text{erfc}$ denotes the complimentary error function, $T_{\infty}$ is the temperature of the cooling fluid and $T_0$ the initial temperature of the solid according to \cite{Ahmed.1998}. The variable $x$ indicates the distance coordinate measured from the cooled front surface of the solid and $t$ is the time since initiation of the cooling process. The parameter $\alpha$ describes the thermal diffusivity of the material, chosen according to $\alpha=10\ \frac{mm^2}{s}$ as an averaged value for Ti-6Al-4V based on \cite{boivineau2006thermophysical}. The parameter $g=h/k$ defines the ratio of the convective heat transfer coefficient $h$ on the surface of the solid to its thermal conductivity $k$. Since the specific heat transfer coefficient $h$ from the experiment is unknown, we determine the parameter $g$ in the following in an inverse manner to optimally match four exemplary cooling curves provided in~\cite{Ahmed.1998}. We relax the original assumption of $g=const.$ underlying the analytic solution \eqref{eqn: semi_infinite_cooling} slightly by introducing a quadratic temperature-dependence for $g(T)$. As discussed above, to make the numerically predicted microstructure evolution comparable to the experiments, the model requires cooling curves that match the cooling behavior from the experiments in good approximation. Considering a potentially temperature-dependent parameter $g(T)$ allows us to represent the experimental cooling curves with higher accuracy. Also from a physical point of view it seems reasonable that  $g(T)$ is not necessarily constant across the large temperature spans relevant for these cooling/quenching experiments:
\begin{equation}
    \label{eqn: quadratic_heuristic}
    g(T) = s_g \cdot \left(a_g + b_g\cdot \frac{T-T_\infty}{T_\infty} + c_g \cdot \left(\frac{T-T_\infty}{T_\infty}\right)^2\right)
\end{equation}
Based on the experimental set-up as described in \cite{Ahmed.1998}, we can directly set $T_0=1323\ \text{K}$. Furthermore, as the continuous cooling of the Ti-6Al-4V probes was described to be conducted with air or water \cite{Ahmed.1998}, we set $T_0=293.15\ \text{K}$, assuming room temperature for the cooling fluid. The cooling rates in experiments can typically be adjusted by either different cooling air speeds or switching to water as a cooling medium for the extreme case of quenching. In order to keep the number of unknown parameters limited, we only considered air cooling in the numerical realization of the cooling curves. To vary the cooling rates as required for the CCT-experiments in the next section, we will adjust the cooling rate by means of an additional scaling parameter $s_g$, which can be interpreted as a manipulation of the cooling air flow velocity. For the inverse parameter identification in this section, we fix the scaling parameter to the default value $s_g=1$. The remaining parameters $\boldsymbol{\theta}_{\text{thermo}}= [a_g, b_g, c_g]^T$ in~\eqref{eqn: quadratic_heuristic} are specific to the performed experiment and are inversely determined based on four temperature curves (a-, b-, c- and d-curve in Figure~\ref{fig: semi_inf_cooling}) experimentally measured at four different positions $x$ along the bar axis at otherwise identical cooling conditions~\cite{Ahmed.1998}.
We use again the maximum likelihood (ML) point estimate $\boldsymbol{\theta}^{*}_{\text{thermo}}$ as an optimal choice for the identified parameters required in Equations \eqref{eqn: semi_infinite_cooling} and \eqref{eqn: quadratic_heuristic}. Under the assumption of Gaussian measurement noise, the ML point estimate is found to be the minimizer of the squared loss function over all temperature curves:
\begin{align}
    \label{eqn: sum_squares}
    \begin{split}
    \boldsymbol{\theta}^*_{\text{thermo}} &= \underset{\boldsymbol{\theta}_{\text{thermo}}}{\mathrm{argmin}}\bigg[\int\limits \left(T_{\text{exp.,a}}(t)-T_{\text{sib}}(x=3.2,t, \boldsymbol{\theta}_{\text{thermo}})\right)^2 dt + \int\limits \left(T_{\text{exp.,b}}(t)-T_{\text{sib}}(x=9.5,t, \boldsymbol{\theta}_{\text{thermo}})\right)^2 dt\\
    &\qquad\qquad+ \int\limits \left(T_{\text{exp.,c}}(t)-T_{\text{sib}}(x=12,t,\boldsymbol{\theta}_{\text{thermo}})\right)^2 dt + \int\limits \left(T_{\text{exp.,d}}(t)-T_{\text{sib}}(x=15.2,t, \boldsymbol{\theta}_{\text{thermo}})\right)^2 dt \bigg]
    \end{split}
\end{align}
In Equation \eqref{eqn: sum_squares}, $T_{\text{exp.,a}}(t),T_{\text{exp.,b}}(t),T_{\text{exp.,c}}(t),T_{\text{exp.,d}}(t)$ refer to the experimental temperature measurements in \cite{Ahmed.1998} and $T_{\text{sib}}(x=3.2,t, \boldsymbol{\theta}_{\text{thermo}}),T_{\text{sib}}(x=9.5,t, \boldsymbol{\theta}_{\text{thermo}}),T_{\text{sib}}(x=12,t, \boldsymbol{\theta}_{\text{thermo}}),T_{\text{sib}}(x=15.2,t, \boldsymbol{\theta}_{\text{thermo}})$ are the corresponding simulated temperature profiles using Equations \eqref{eqn: semi_infinite_cooling} and \eqref{eqn: quadratic_heuristic}. It is emphasized that all four experimental and corresponding model-based temperature curves (differing only in the measurement/evaluation position $x$) are considered simultaneously in the inverse analysis. The broader data basis resulting from this combined approach is a pre-requisite for a robust inverse identification procedure and an accurate representation of the experimental cooling curves. We utilize again the Levenberg-Marquardt optimizer \cite{more1978levenberg} to solve for $\boldsymbol{\theta}^*_{\text{thermo}}$ in Equation \eqref{eqn: sum_squares}, which results in the identified parameter set ${\boldsymbol{\theta}^*_{\text{thermo}}=[a_g, b_g, c_g]^T=[73.8, -39.3, 6.3]^T \ \text{m}^{-1}}$. Figure \ref{fig: semi_inf_cooling} shows a good agreement of the analytic model with the temperature measurements by Ahmed et al. \cite{Ahmed.1998}. Equation \eqref{eqn: semi_infinite_cooling} based on the identified parameters can now be used to generate arbitrary cooling curves for the simulation of CCT-experiments by varying the scaling parameter $s_g$ in \eqref{eqn: quadratic_heuristic}, as shown in the following section.

\subsubsection{Simulation-based prediction of CCT-diagrams for Ti-6Al-4V}
\begin{figure}[htbp]
        \centering
        \includegraphics[scale=0.35]{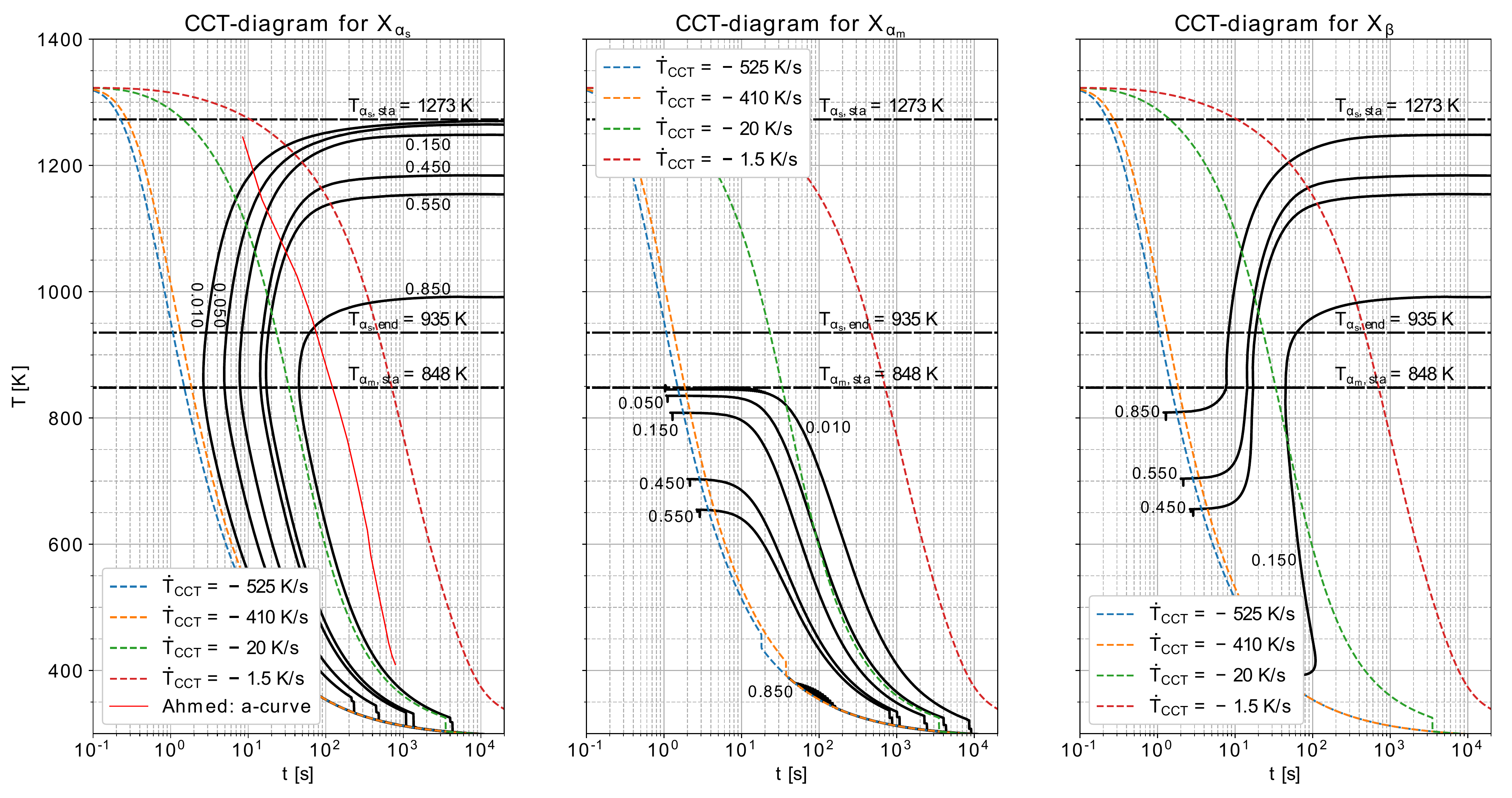}
         \caption{CCT-diagram for the $\as$-, $\am$- and $\beta$-phases resulting from the proposed microstructure model for the identified parameter set $\boldsymbol{\theta}^*_{\text{thermo}}$ along with four characteristic cooling curves according to~\eqref{eqn: semi_infinite_cooling} and the experimental a-curve from~\cite{Ahmed.1998}. Left: contour-lines for $\Xa$. Middle: contour-lines for $\Xm$. Right: contour lines for $\Xb$.} 
        \label{fig: CCT_diagram}
    \end{figure}

For the simulation-based creation of CCT-diagrams, we chose an equidistant range of scaling parameter values $s_g$ such that 150 cooling curves between $\dot{T}_{\text{CCT}}=-1 \ K/s$  and $\dot{T}_{\text{CCT}}=-600 \ K/s$ were realized. Afterwards, the evolving microstructure was again recorded over the $t\times T$-coordinate space in form of contour-lines for the individual phase fractions. Figure \ref{fig: CCT_diagram} displays the results of the CCT-simulations, along with four characteristic cooling curves analytically reproduced by means of~\eqref{eqn: semi_infinite_cooling} as well as the aforementioned experimental a-curve from~\cite{Ahmed.1998}. 
Note that in contrast to the TTT-diagrams in Figure \ref{fig: TTT_diagram}, the isolines in Figure~\ref{fig: CCT_diagram} show the true/absolute phase fractions such that the phase fraction values of all three plots would sum up to one for any given data point. In \cite{Ahmed.1998} it was found that cooling rates faster than $\dot{T}_{\text{CCT}}=-410 \ K/s$ resulted in a fully martensitic transformation of the microstructure while cooling rates below $\dot{T}_{\text{CCT}}=-20 \ K/s$ led to a fully diffusional $\as$-formation. According to Figure \ref{fig: CCT_diagram}, the results predicted by our model  are in very good agreement with these findings as the contour-line for $\Xa=1\%$ coincides almost exactly with the cooling line of $\dot{T}_{\text{CCT}}=-410 \ K/s$ (see dashed orange line in Figure \ref{fig: CCT_diagram}, left) meaning that the proposed model predicts a fully martensitic transformation for cooling rates faster than $\dot{T}_{\text{CCT}}=-410 \ K/s$. We want to emphasize that the inverse identification of the microstructure model parameters was solely based on the previously presented TTT-data and not on the CCT data. Thus, the accurate prediction of the CCT diagram confirms the validity of our microstructure model. In contrast to existing approaches, the proposed microstructure model does not explicitly enforce this critical cooling rate as transformation criteria, instead a lower formation bound at roughly $\dT=-410\ K/s$ emerges naturally form the dynamics of the diffusion equations, which is a very strong argument for the generality and consistency of the proposed modeling approach. Furthermore, the predicted results show also a fully diffusional $\as$-formation (i.e. $\Xm$ is close to zero) for cooling rates slower than $\dot{T}_{\text{CCT}}=-20 \ K/s$ (see dashed green line in Figure \ref{fig: CCT_diagram}), which is also in very good agreement with the experimental observations by Ahmed et al. \cite{Ahmed.1998}. Besides these characteristic quantitative values, also the overall qualitative appearance of the predicted CCT-diagrams matches theoretical presentations in \cite{cahn1956transformation, Ahmed.1998} very well.

\FloatBarrier
\section{Part-scale demonstration: Microstructure evolution for an exemplary SLM process}
\label{sec: part_scale_demonstration}

In the context of SLM process simulation, macroscale thermo-mechanical models are typically applied to predict the temperature evolution, residual stresses and dimensional warping~\cite{Cervera1999,Childs2005,Zaeh2010, Shen2012,Hodge.2014,Denlinger2014,Hodge.2016,Kollmannsberger2017,Riedlbauer2017,Roy2018,Bartel2018,Zhang2018a,Kollmannsberger2019,Neiva2019,Proell2020,Noll2020}. In the following, an advanced macroscale SLM model of this type~\cite{Hodge.2014, Hodge.2016} will be applied to provide the temperature field during SLM as required by the proposed microstructure model. We will first introduce the (thermal part of the) SLM model and the numerical set-up, then demonstrate the microstructure evolution during SLM for selected points of the domain over time and eventually show snapshots of micro-structural states for cross-sections and different base-plate temperatures.
\subsection{Macroscale thermal model for SLM process}
\label{sec: effective_medium}

The microstructure model is implemented in the parallel, implicit
finite element code \emph{Diablo} \cite{diablo_manual_2018}.  The evolution of the microstructure
is  driven by a one way coupling of the emerging thermal field. The latter is the solution of the balance of thermal energy, and the associated boundary and initial conditions:
\begin{subequations}
\begin{align}
\rho c_p \dot{T} = -\operatorname{div} \mathbf{q} + r,& \; \; \mbox{ in } \Omega,\label{eq:balance_thermal_energy}\\
T\left(\mathbf{x}_T,t\right) = \bar{T},& \; \; \mbox{ for } \mathbf{x}_T \in \Gamma_T\\
q\left(\mathbf{x}_q,t\right) = \bar{\mathbf{q}} \cdot \mathbf{n},& \; \; \mbox{ for } \mathbf{x}_q \in \Gamma_q\\
T\left(\mathbf{x},0\right) = T_0,& \; \; \mbox{ on } \Omega \cup \partial \Omega,
\end{align}
\end{subequations}
where~$\Gamma_T$ is the portion of the total boundary~$\partial \Omega$ subject
to essential boundary conditions, and~$\Gamma_q$ is the portion of the total
boundary subject to natural boundary conditions.  The constitutive behavior
is characterized by a temperature-dependent Fourier conduction of the form
\begin{align}
\label{eqn: fourier_cond}
\mathbf{q} = -\mathbf{k} \operatorname{grad} T,
\end{align}
where~$\mathbf{k}$ is a second-order tensor of thermal conductivities, which may
also be a function of spatial coordinates. For the current problem, the heat source~$r$ in Equation
(\ref{eq:balance_thermal_energy}) plays an important role, that being to represent
the deposition of laser energy into a powder.  While there exist various models in
the literature, the model in the current implementation
was taken from \cite{gusarov_heat_2007} in an effort to
provide the most natural description of the physical process.

The numerical implementation consists of an iterative nonlinear solver that uses
consistent Fr\'{e}chet derivatives, with the solution computed on first-order finite
elements via an iterative method to solve the linear system of equations.  Time
integration is performed using the generalized trapezoidal rule.  The code uses
distributed memory parallelism to speed up the solution of the spatial problem.
The solution of the thermal field is evaluated at the Gauss points of the finite element discretization. For a full description of the model and its implementation, see \cite{Hodge.2014, Hodge.2016}. While \emph{Diablo} contains additional physics, including solid mechanics and mass transfer, a detailed description is omitted at this point and we direct the interested reader to \cite{diablo_manual_2018}.

\begin{figure}[htbp]
\begin{subfigure}[t]{\textwidth}
\centering
\includegraphics[scale=.19]{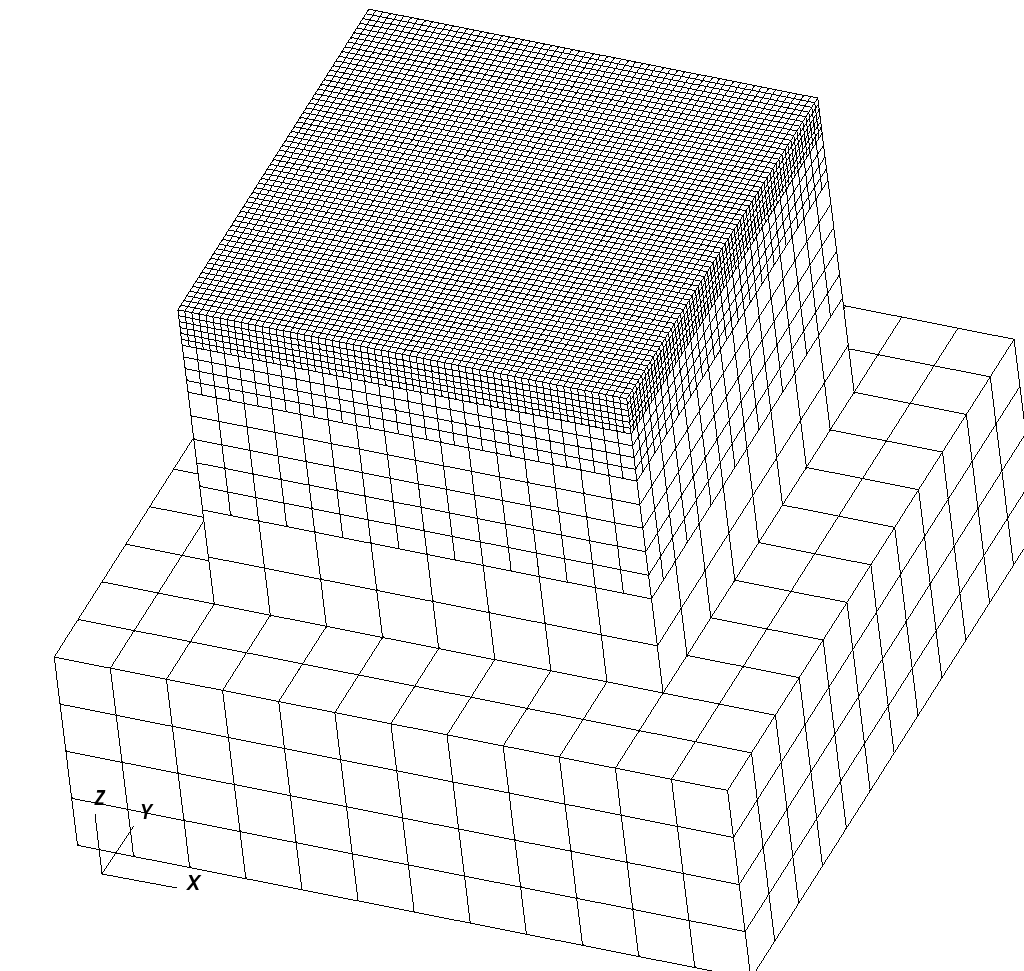}
\end{subfigure}

\begin{subfigure}[t]{\textwidth}
\centering
\includegraphics[scale=.26]{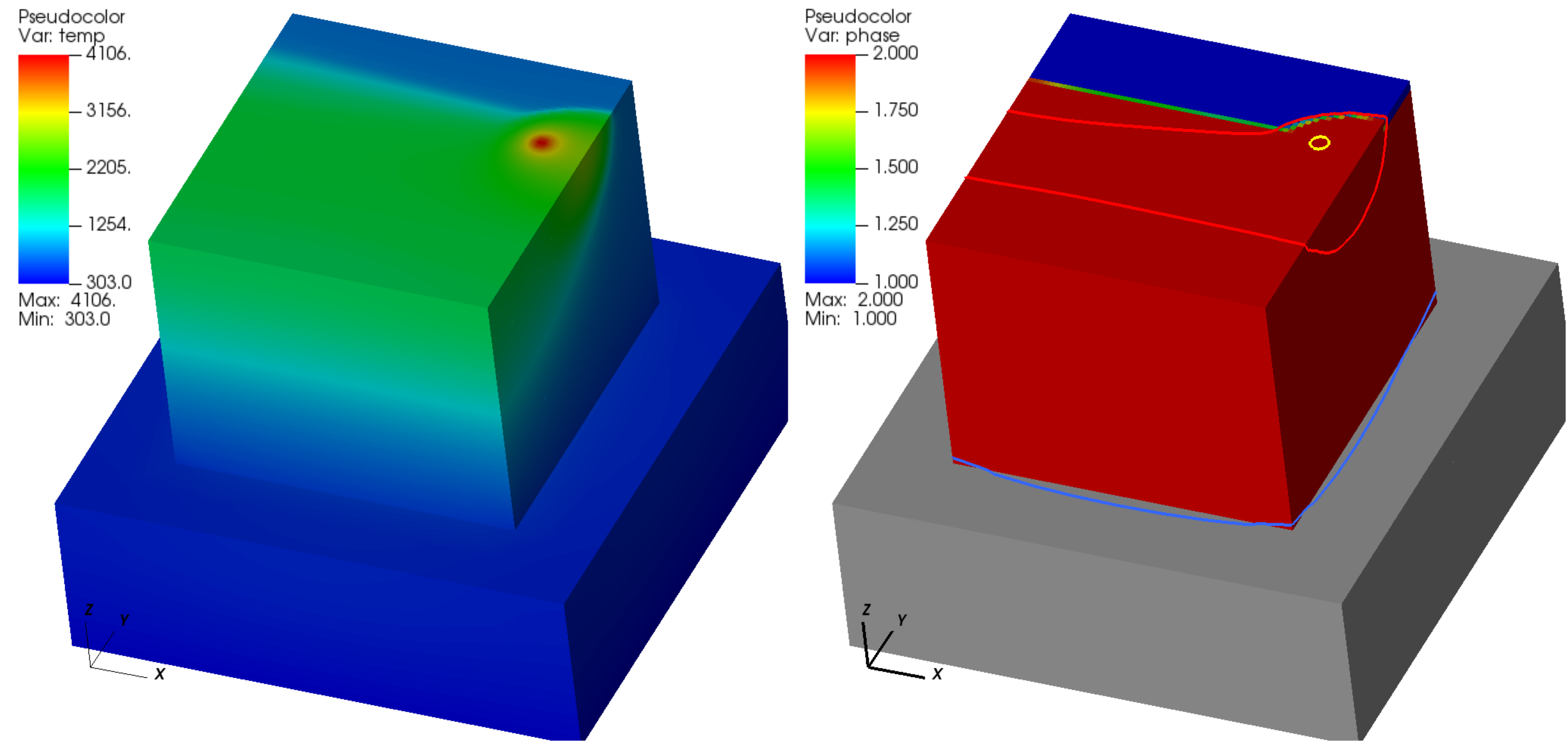}
\end{subfigure}

  \caption{Representative state of the thermal model.  The top frame depicts the mesh, for which refinement is mediated by distance from the free surface, the bottom left frame depicts the temperature field, and the bottom right frame depicts the internal phase variable (with gray representing the baseplate, which is consolidated material throughout the simulation) as well as containing several temperature contours, with blue being $T=505\ K$, red being the melt temperature, and yellow being the vaporization temperature.} 
  \label{fig: 3D_cube}
 \end{figure}

\FloatBarrier
In the numerical demonstration, we are investigating the selective laser melting process of a one-millimeter-sided cube onto a base-plate subject to different values of constant Dirichlet boundary conditions ${T_{\text{bp}}\in\{303\ K,\ 500\ K,\ 700\ K,\ 900\ K,\ 1100\ K,\ 1300\ K\}}$ (which are applied solely on the ``bottom'' of the baseplate, $\Gamma_T=\min\left(z\right)$) and the associated effect on the emerging microstructure distribution.  The domain is mostly insulated, with the exception of the free surface ($\Gamma_q=\max\left(z\right)$), for which there exists a Neumann boundary condition that accounts for the energy loss due to both radiation and evaporation.  The initial condition is defined as $T_0=303\,K$.  The processing parameters consist of a laser travelling at $600\, mm/s$, with a power of $100\, W$, beam radius of $30\, \mu m$, and track spacing of $120\, \mu m$, with eight laser tracks per layer. The scanning direction is in each subsequent layer rotated by 45 degrees counterclockwise. The powder layer depth is $30\, \mu m$, such that the final cube consists of 34 layers. The physical time of investigation is $t\in [0\ s, 15\ s]$, with the active processing time occurring over the interval $\left[0\, s,\, 3.58\, s\right]$. The geometric set-up and one snapshot of the emerging temperature field and phase variable are depicted in Figure~\ref{fig: 3D_cube}.

These problems were run using 128 cores, and typically took around 15 to 20 hours of wall clock time to complete.  The significant resources needed to simulate laser powder bed fusion (LPBF) problems depend on several factors, including its multiscale nature, which can be relevant with respect to both the spatial and temporal domains.  The spatial scales are resolved via the use of $h$-refinement, with refinement indicators tailored to the LPBF problem, as can be seen in Figure~\ref{fig: 3D_cube}.  With respect to the temporal scales, is it noted that the large\footnote{As an example, a cube with $2\, cm$ edges, run with fully nonlinear behavior and physically relevant process parameters, requires around 500 million time steps.} number of time steps (associated with the active simulation time and the high speed of the laser) necessitates more sophisticated methodological treatment in order to reduce the wall clock run times to the useful range without significantly degrading the solution accuracy.  Both the geometric and temporal scale issues are currently being addressed by the \emph{Diablo} development team, as addressed in the manuscripts \cite{ganeriwala_towards_2020, hodge_towards_2020}.

\FloatBarrier
\subsection{Microstructure evolution for the center node of the one millimeter-sided cube}
We start the part-scale demonstrations by analyzing the microstructure evolution over time at the cube's center point ${P=[0.0,0.0,0.5]^T \ mm}$. The layer that contains this point is processed at $t\approx1.72\ s$. Figure \ref{fig: center_303} to \ref{fig: center_1300} depict the respective temperature profiles along with the temperature rates and the corresponding microstructure evolution for all six investigated base-plate temperatures. The temperature profile is supported by characteristic temperatures such as the Martensite-start temperature $\Tms$, the $\as$-end and start temperatures, $\Tbs$ and $\Tbe$ as well as the solidus temperature $\Ts$ and liquidus temperature $\Tl$. 

For $T_{\text{bp}}<900\ K$, respectively in Figure \ref{fig: center_303} to \ref{fig: center_700}, we can observe the formation of Martensite directly after the first melting by the laser in the time period $1.72\ s<t<3\ s$ and then again slightly before $t=4\ s$, followed by a subsequent stabilization of the Martensite phase

The amount of formed Martensite phase for $t>4 \ s$ is dependent on the metastable Martensite equilibrium phase fraction$\Xmeq$ in Equation \eqref{eqn: am_eq_general} and decreases with rising $T_{\text{bp}}$. As stated in Equation \eqref{eqn: alpha_total_equ_new}, the metastable Martensite phase will transform to $\as$ for $t\rightarrow\infty$. Nevertheless, the diffusion rate for $\Xm$ transformation is at lower temperatures ($T<700 \ K$) so small (see Figure \ref{fig: k_rate_T}) that Martensite dissolution, respectively the formation of stable $\as$ remains unnoticeable.

At a base-plate temperature of $T_{\text{bp}}=900\ K$ (see Figure \ref{fig: center_900}) no Martensite formation is recorded as the center node's temperature stays above the Martensite start temperature $\Tms$ but below the $\alpha$-transus end temperature such that $\Tms<T_{\text{bp}}<\Tbs$. The latter condition will result in the maximum amount of stable $\as$-transformation as the $\alpha$-equilibrium phase fraction $\Xaeq=0.9$ at this temperature is still the same as at room temperature. Figure \ref{fig: center_900} only shows the first 13 seconds of the SLM process but the $\as$-nucleation will continue until $\Xa=\Xaeq=0.9$. Other points within the cube might nevertheless experience a martensitic transformation when the steady-state temperatures fall below the Martensite start temperature $\Tms$. 

For a base-plate temperature of $T_{\text{bp}}=1100 \ K$ (see Figure \ref{fig: center_1100}), which fulfills $\Tms<\Tbs<T_{\text{bp}}<\Tbe$, we observe a similar characteristic but at slower $\as$-formation rates and a lower equilibrium phase fraction $\Xaeq<0.9$.

Eventually, for a base-plate temperature of $T_{\text{bp}}=1300\ K$ (Figure \ref{fig: center_1300}), which is above $\Tbe$, the microstructure at $P$ stays in the $\beta$-regime for all times (as long as the base-plate temperature is held at $T_{\text{bp}}=1300\ K$) and no $\as$- or $\am$-transformation is observed.

We can summarize the following points for the microstructure evolution at the cube's center node $P$:
\begin{itemize}
\item Without preheating of the base-plate, respectively for moderate preheating ($T_{\text{bp}}<\Tms$), the Martensite phase is dominating the microstructure in the deposited material.
\item For $T_{\text{bp}}\le 900 \ K$ the center point experiences six liquid-solid transitions that are  followed by martensitic transformations. For higher base-plate temperatures $T_{\text{bp}}>\Tms$ the material melts more often but the martensitic transformation is not triggered anymore. The visible temperature peaks in Figures \ref{fig: center_303} to \ref{fig: center_1300} consist actually of several very closely packed peaks that are caused by the neighboring laser tracks that heat up the center node $P$ again. This process happens so fast (the scanning speed in the simulation was $600\ mm/s$) that these peaks are not distinguishable on the presented time-scale. The 17 visible peaks corresponds to the number of layers that are processed above the center node $P$ such that the cube consists in total of 34 layers.
\item The absence of martensitic transformation for $T_{\text{bp}}\ge \Tms$ in combination with  slower formation dynamics of the stable $\as$-phase is especially interesting from a mechanical perspective. The $\beta$-phase is known to be softer than the $\as$- or $\am$-phase \cite{Warwick.2012}, such that build-up of residual stresses is expected to be smaller if the microstructure is dominated by the $\beta$-phase for longer times.
\item Theoretically, Martensite will dissolve for $t\rightarrow\infty$ but the transformation will only be noticeable for ${T>700 \ K}$.
\end{itemize}

\begin{figure} 
\centering
    \begin{subfigure}[htbp]{1.\textwidth}
    \includegraphics[width=1\linewidth]{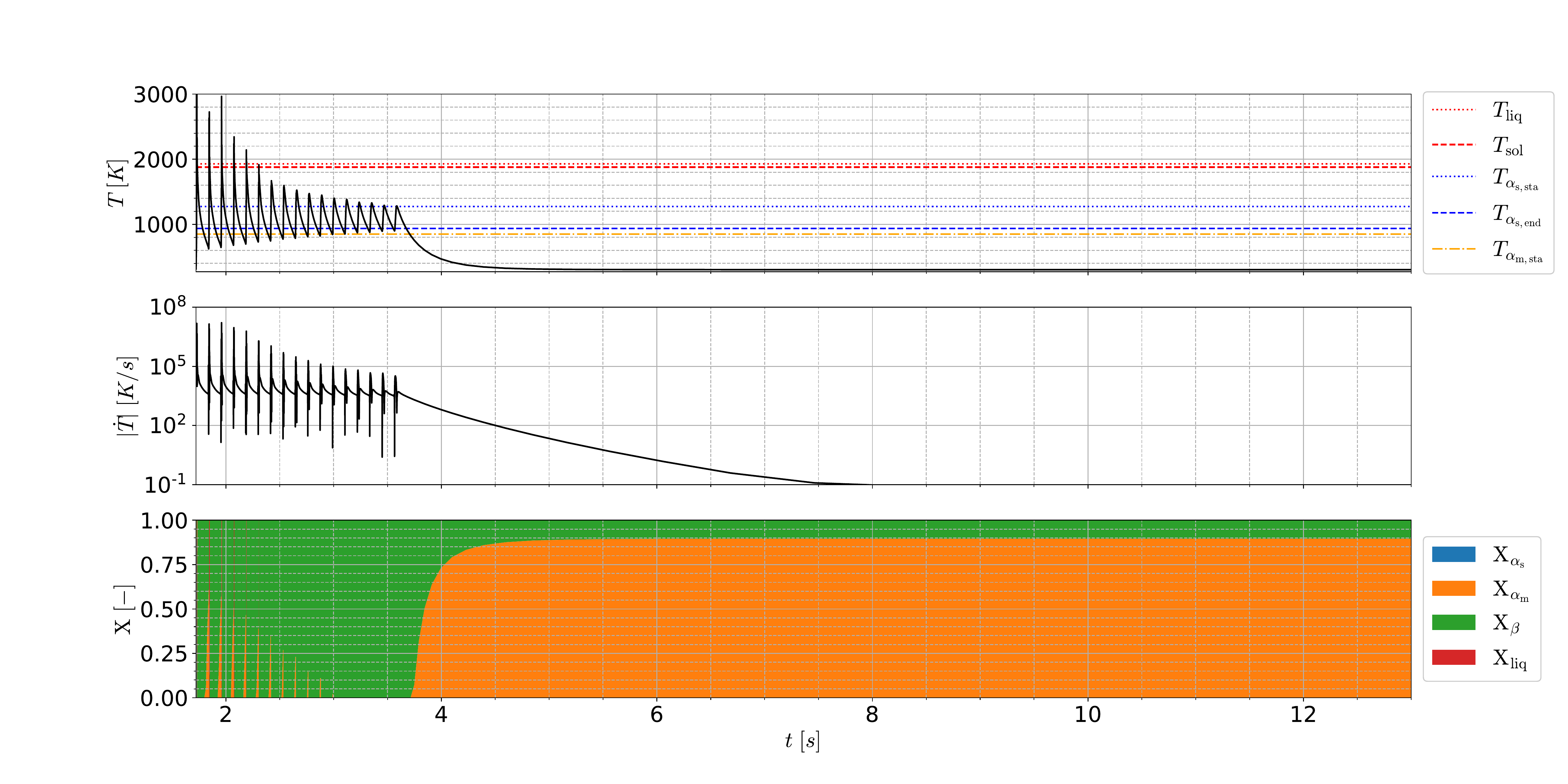}
    \newsubcap{Simulated microstructure evolutions during the SLM process of a one millimeter-sided cube at its center point $P$ which is processed at $t\approx1.72\ s$ with base-plate temperature $T_{\text{bp}}=303 \ K$.}
    \label{fig: center_303}
    \end{subfigure}
    
\end{figure}

 \begin{figure}    
 \ContinuedFloat  
\centering   
    \begin{subfigure}[b]{1.\textwidth}
    \includegraphics[width=1\linewidth]{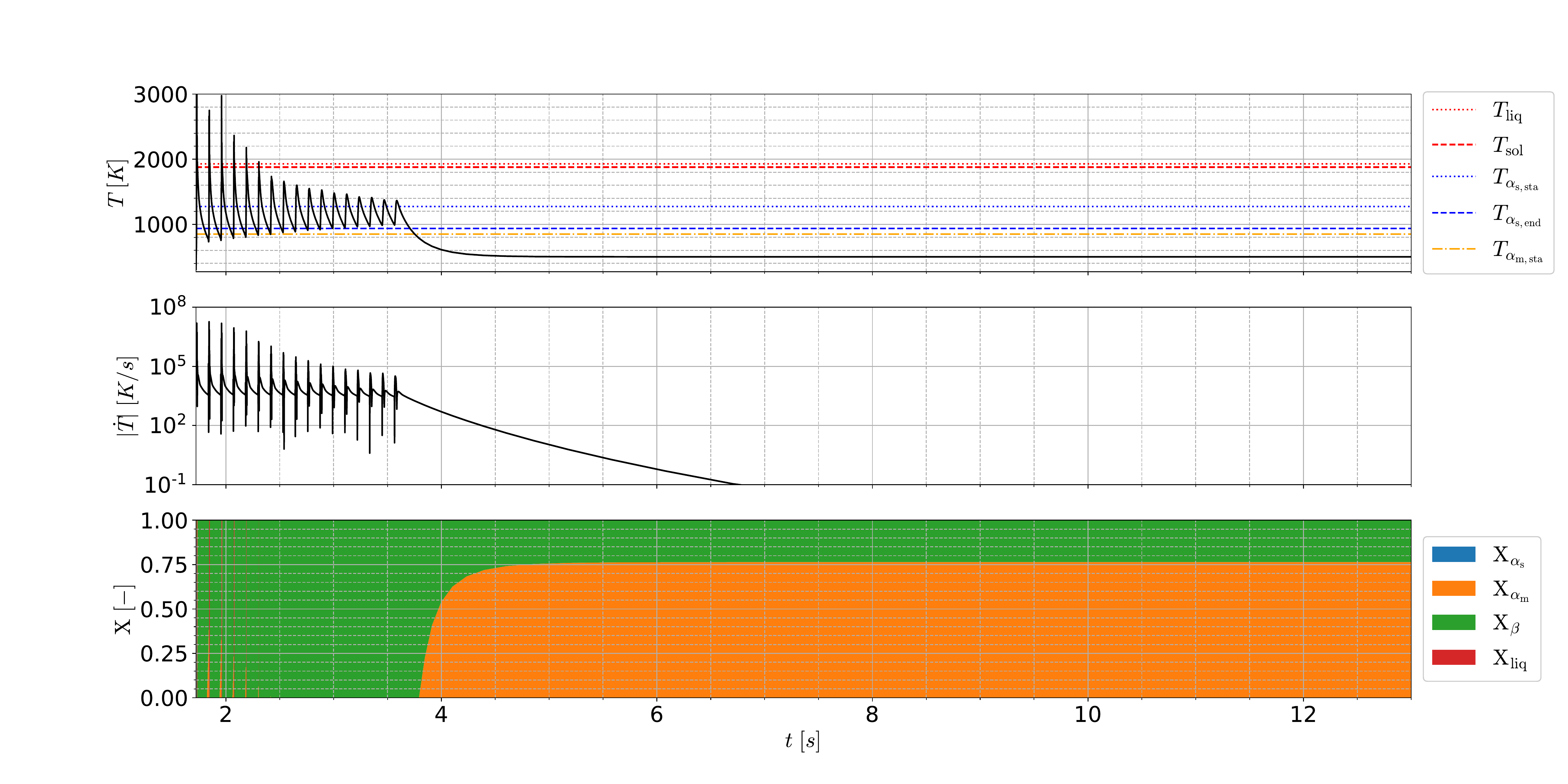}
    \newsubcap{Simulated microstructure evolutions during the SLM process of a one millimeter-sided cube at its center point $P$ which is processed at $t\approx1.72\ s$ with base-plate temperature $T_{\text{bp}}=500 \ K$.}
    \label{fig: center_500}
    \end{subfigure}
   
    \begin{subfigure}[b]{1.\textwidth}
    \includegraphics[width=1\linewidth]{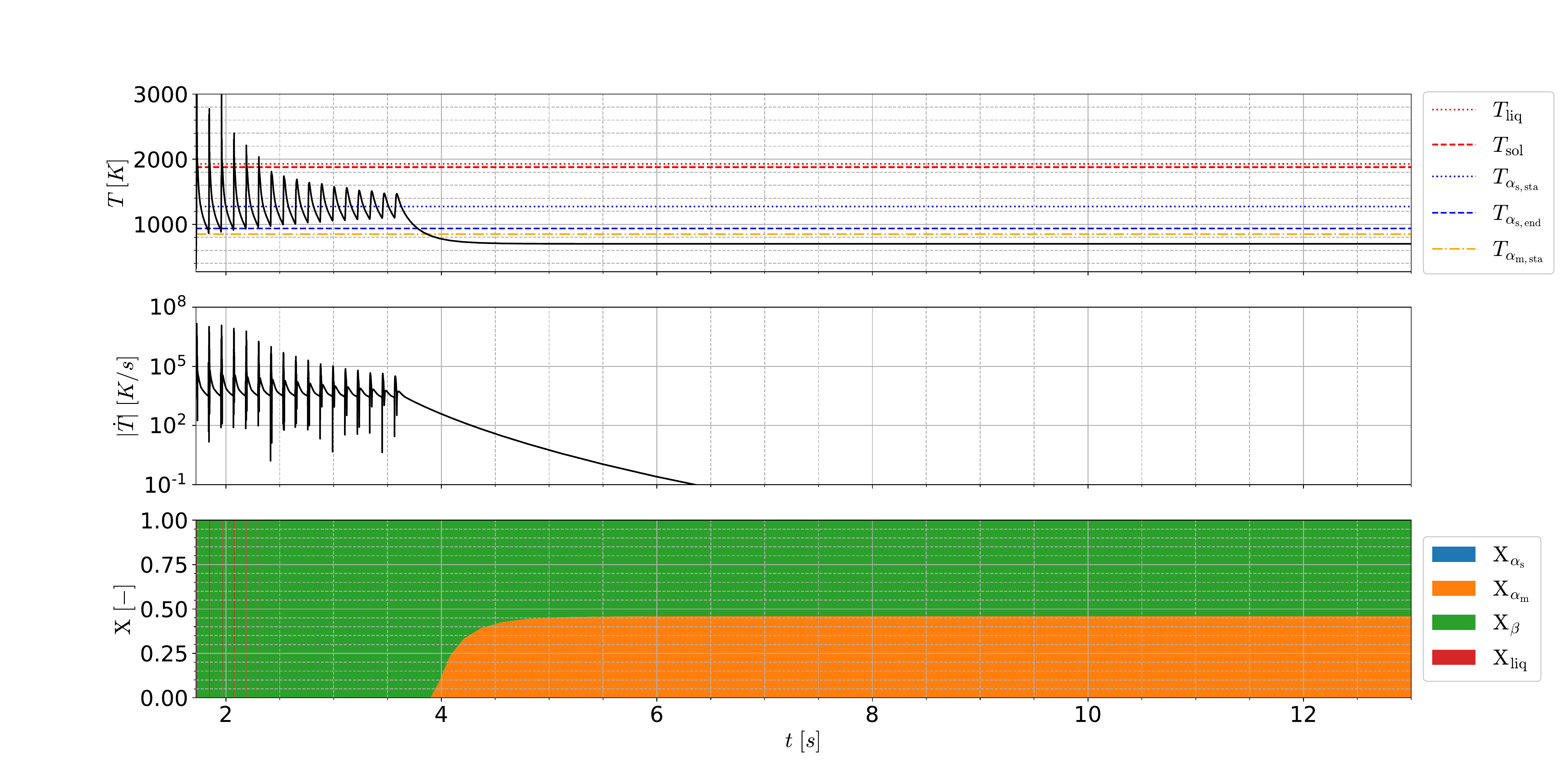}
    \newsubcap{Simulated microstructure evolutions during the SLM process of a one millimeter-sided cube at its center point $P$ which is processed at $t\approx1.72\ s$ with base-plate temperature $T_{\text{bp}}=700 \ K$.}
    \label{fig: center_700}
    \end{subfigure}
\end{figure}
    
 \begin{figure} 
\ContinuedFloat   
    \begin{subfigure}[b]{1.\textwidth}
    \includegraphics[width=1\linewidth]{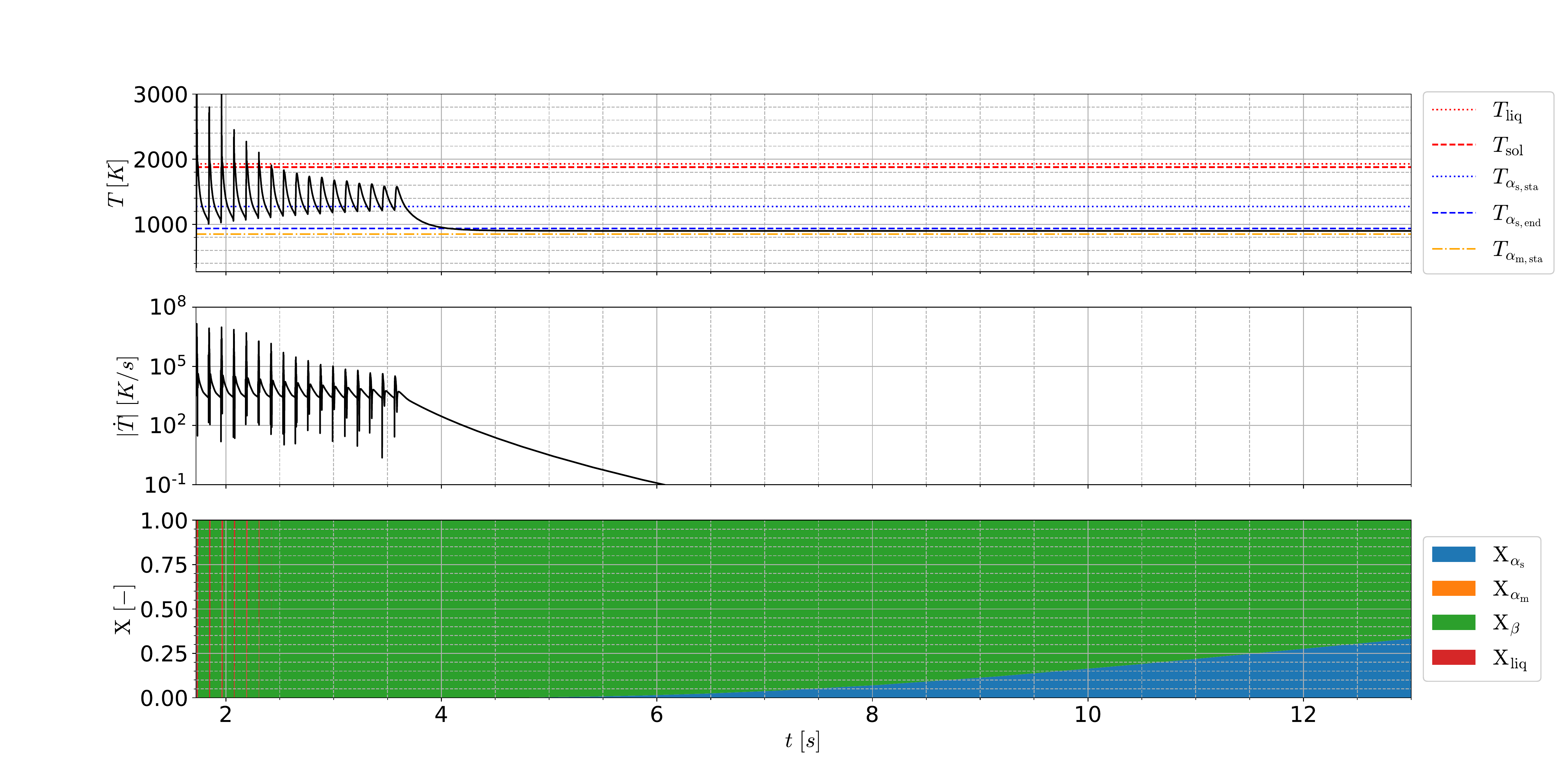}
    \newsubcap{Simulated microstructure evolutions during the SLM process of a one millimeter-sided cube at its center point $P$ which is processed at $t\approx1.72\ s$ with base-plate temperature $T_{\text{bp}}=900 \ K$.}
    \label{fig: center_900}
    \end{subfigure}
    
    \begin{subfigure}[b]{1.\textwidth}
    \includegraphics[width=1\linewidth]{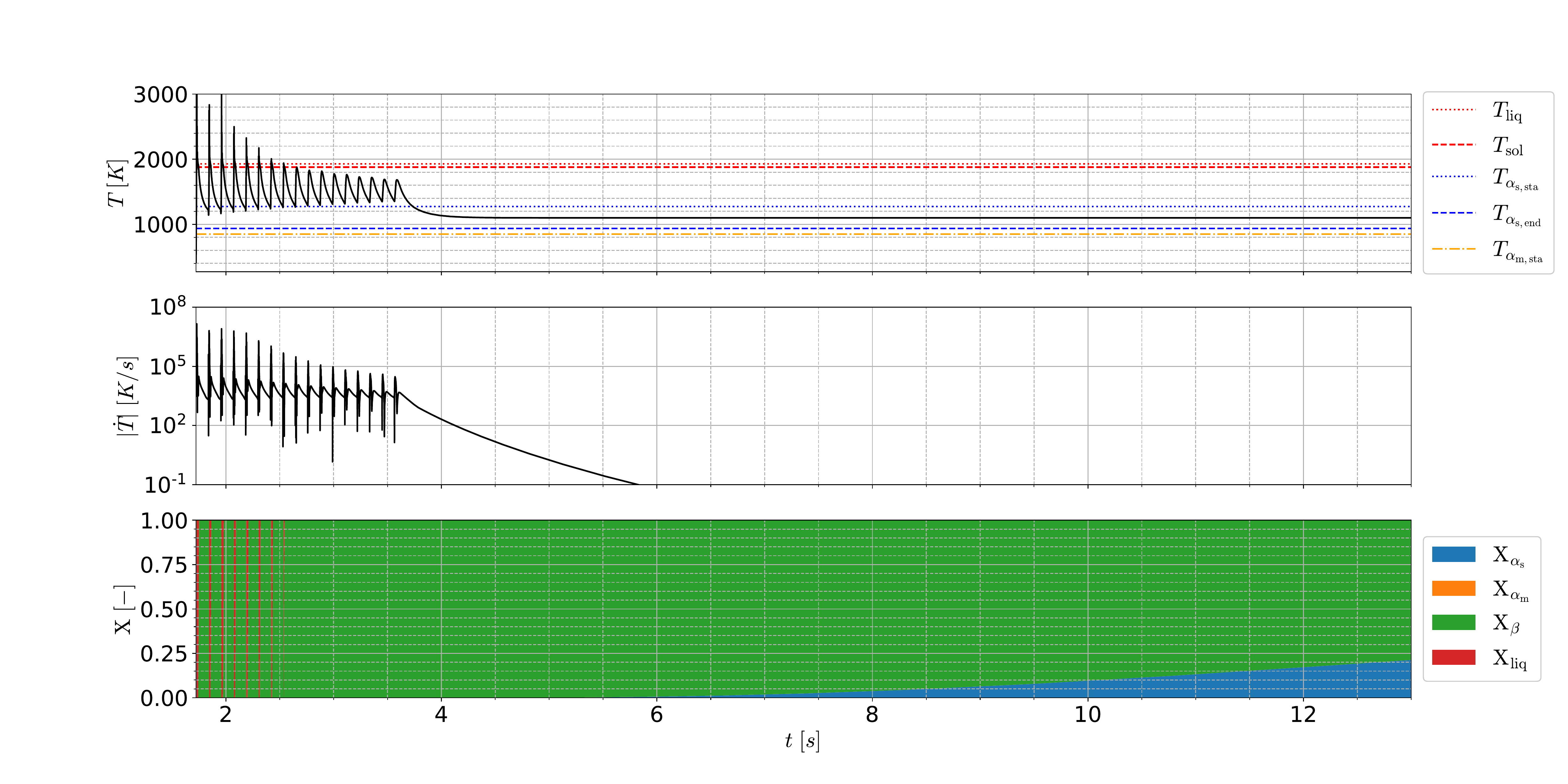}
    \newsubcap{Simulated microstructure evolutions during the SLM process of a one millimeter-sided cube at its center point $P$ which is processed at $t\approx1.72\ s$ with base-plate temperature $T_{\text{bp}}=1100 \ K$.}
    \label{fig: center_1100}
    \end{subfigure}
    
\end{figure}

\begin{figure}
\ContinuedFloat   
    \begin{subfigure}[b]{1.\textwidth}
    \includegraphics[width=1\linewidth]{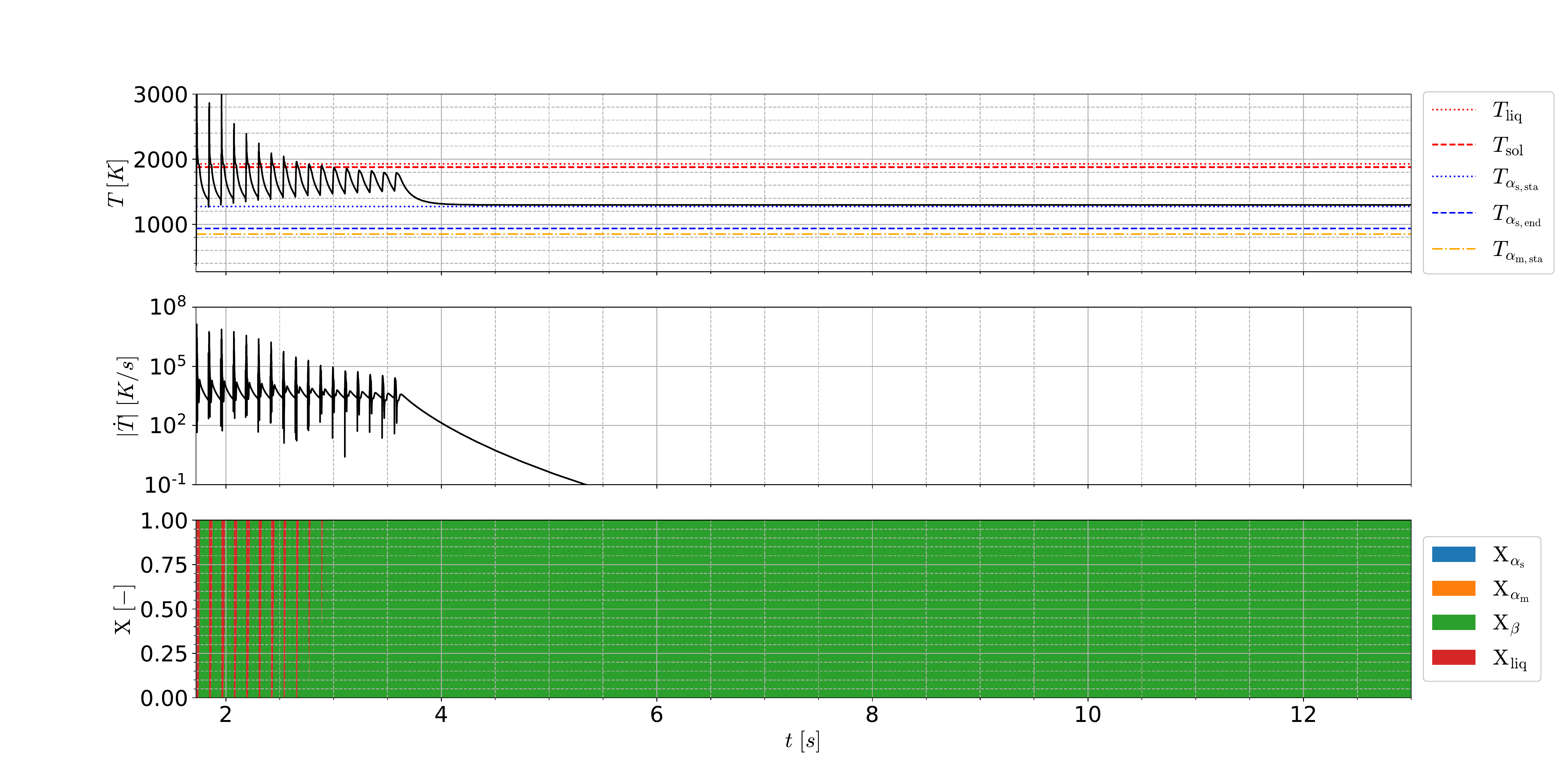}
    \newsubcap{Simulated microstructure evolutions during the SLM process of a one millimeter-sided cube at its center point $P$ which is processed at $t\approx1.72\ s$ with base-plate temperature $T_{\text{bp}}=1300 \ K$.}
    \label{fig: center_1300}
    \end{subfigure}
    
\caption{Simulated microstructure evolutions during the SLM process of a one millimeter-sided cube at its center point $P$ which is processed at $t\approx1.72\ s$. The sub-figures show the evolving microstructure for different base-plate temperatures $T_{\text{bp}}\in\{303,500,700,900,1100,1300\}\ K $.}
\end{figure}
 
\newpage
\subsection{Microstructure distributions in center plane of the one millimeter-sided cube}
\label{sec: slm_cross_sec}
In this section, we continue the investigations of the evolving microstructure in the SLM example of the one millimeter sided cube. In contrast to the former local analysis of the center node, we now want to investigate the resulting microstucture distributions within a complete cross-section in the vertical x-z-center plane that contains the center node $P$ of the cube, for several selected snap-shots in time. We furthermore consider the exemplary base-plate temperatures $T_{\text{bp}}=303 \ K$ and $T_{\text{bp}}=900 \ K$ to compare the room-temperature case with the first preheated case that does not result in martensitic transformation. Snapshots are taken for the physical times $t\in[3,\ 4,\ 15] \ s$ and phase fractions $\Xa, \Xm$ and $\Xb$ are depicted in Figures \ref{fig: plane_303K3s} to \ref{fig: plane_900K15s}.

For a base-plate at room temperature ($T_{\text{bp}}=303\ K$) we observe spatially strongly heterogeneous microstructure distributions during the ongoing SLM process (see Figure \ref{fig: plane_303K3s} and \ref{fig: plane_303K4s}). The Martensite phase in these cases is propagating upwards (in positive z-direction) starting from the base-plate towards the last processed layer.
 
When looking at the results for the two investigated cases at $t =15 \ s$ (see Figure \ref{fig: plane_303K15s} and \ref{fig: plane_900K15s}), we find a rather homogeneous microstructure distribution in both cases with only a slight deviation for the case $T_{\text{bp}}=900\ K$. For $T_{\text{bp}}=300\ K$, we observe an almost complete Martensite transformation. On the contrary, for $T_{\text{bp}}=900\ K$ no Martensite is formed. Note, that for this case the phase transition from $\as$ to $\beta$ is still ongoing (until an equilibrium phase fraction of $\Xa \approx 0.9$ is reached) at the depicted snap shot at $t =15 \ s$. Moreover, we notice a slightly higher $\as$-concentration at the bottom of the cube, i.e. a higher amount of $\beta$ has already transformed into $\as$ at $t =15 \ s$, which is caused by the higher base-plate temperature that leads to higher nucleation rates $k_{\rightarrow \as}(T)$ (see Equation \eqref{eqn: diff_rate_model}). 

Coming back to the case $T_{\text{bp}}=300\ K$, it is emphasized that for larger geometries we would in general expect a more heterogeneous microstructure state due to a more heterogeneous temperature distribution with strong spatial gradients at the surfaces of the geometry. An extended dwelling time at elevated temperatures, due to the increased thermal mass of larger parts in combination with an increasing amount of absorbed laser energy, would result in longer times at a temperature band ($\Tms < T < \Tbe$) favoring the diffusional dissolution of $\am$ into $\as$ in the core region of such parts, while the higher cooling rates and lower temperature levels at the free surfaces are expected to foster remaining Martensite phase fractions. In the next example, a component of larger size will be analyzed to verify these considerations.
 \begin{figure}[htbp]
 \centering
    \begin{subfigure}[b]{1.0\textwidth}
    \includegraphics[width=1\linewidth]{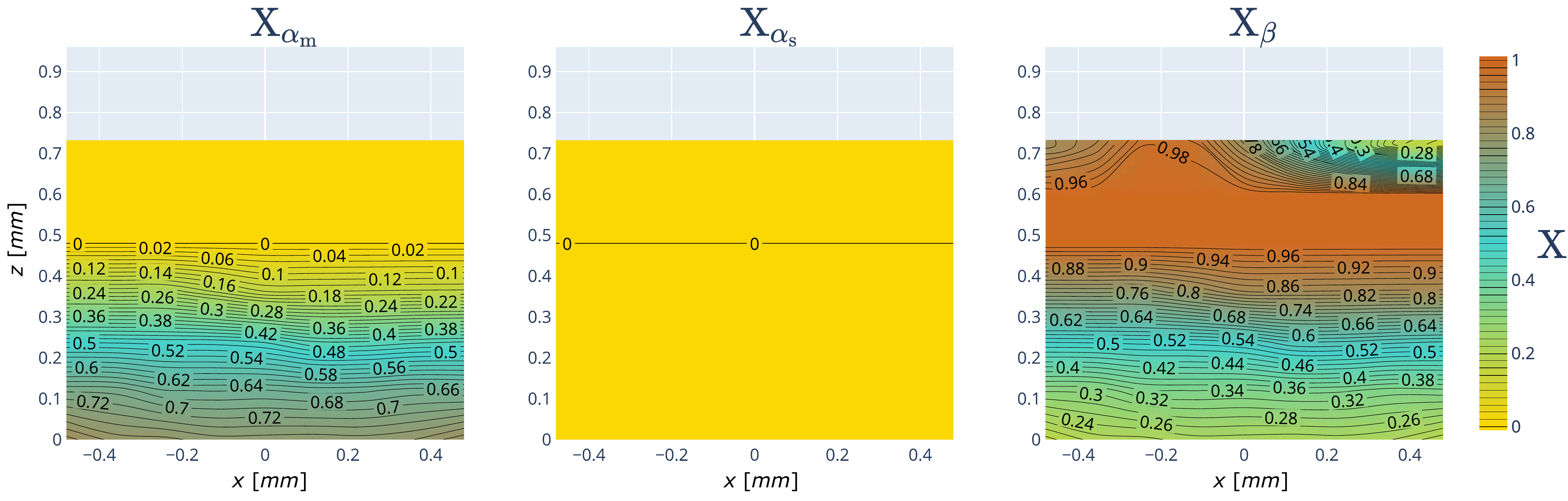}
    \newsubcap{Simulated microstructure distributions for a SLM process of a one millimeter-sided cube within its vertical x-z-center plane that contains the center node $P$. The microstructure state is shown for a base-plate temperature of $T_{\text{bp}}=303 \ K$ and the processing time $t=3 \ s$. The gray area above marks the layers that are not yet processed by the laser at the time $t=3\ s$. In the right figure, which depicts the $\beta$-phase fraction $\Xb$, the melt-pool is indirectly visible through the decreased $\beta$-phase fraction in the right corner of the up-most layers due to the laser that has just previously scanned this plane in x-direction from left to right. In a similar fashion, the decreased amount $\Xb$ in the upper left corner of the right Figure stems from the heat of the laser that is currently melting the subsequent track with a now $45^\circ$ rotated scanning direction (diagonal scanning), starting in a corner of the cube.}
    \label{fig: plane_303K3s}
    \end{subfigure}
    
    \begin{subfigure}[b]{1.0\textwidth}
    \includegraphics[width=1\linewidth]{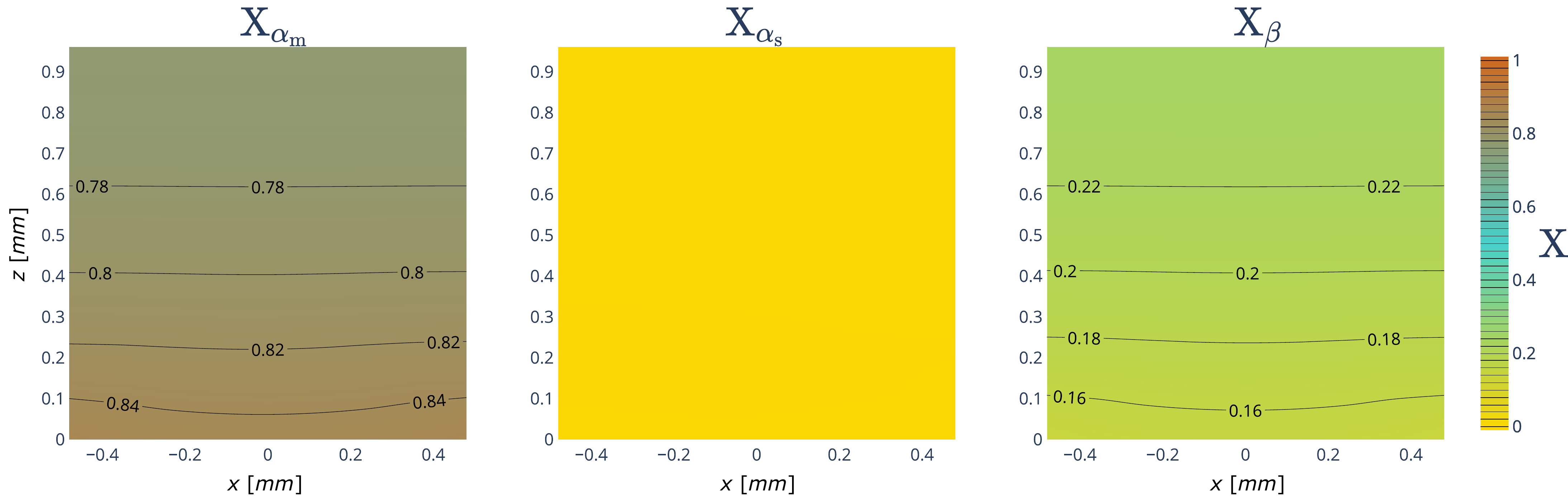}
    \newsubcap{Simulated microstructure distributions for a SLM process of a one millimeter-sided cube within its vertical x-z-center plane that contains the center node $P$. The microstructure state is shown for a base-plate temperature of $T_{\text{bp}}=303 \ K$ and the processing time $t=4 \ s$. Martensitic transformation propagates from the bottom of the cube to its top, which can be seen by the higher $\Xm$-phase fractions towards the bottom of the cube in the left figure.}
    \label{fig: plane_303K4s}
    \end{subfigure}
    
    \begin{subfigure}[b]{1.0\textwidth}
    \includegraphics[width=1\linewidth]{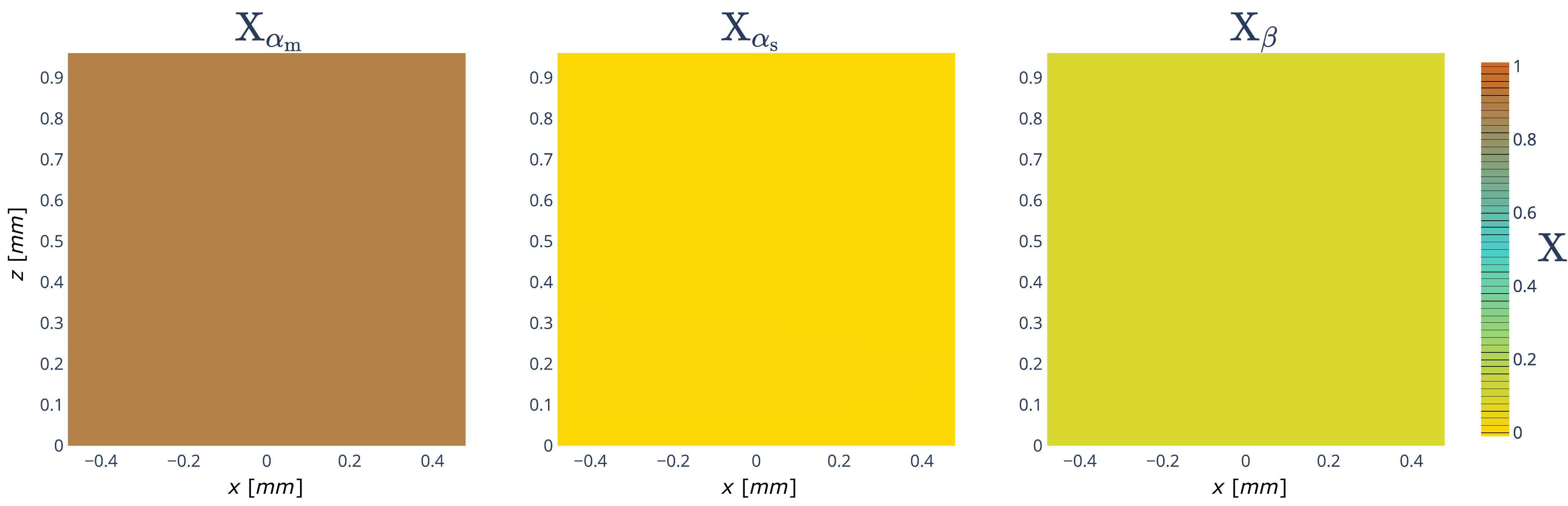}
    \newsubcap{Simulated microstructure distributions for a SLM process of a one millimeter-sided cube within its vertical x-z-center plane that contains the center node $P$. The microstructure state is shown for a base-plate temperature of $T_{\text{bp}}=303 \ K$ and the processing time $t=15 \ s$. A homogeneous microstructure state with full martensitic transformation ($\Xm=0.9$) can be observed throughout the cube.}
    \label{fig: plane_303K15s}
    \end{subfigure}
 \end{figure}
 
\begin{figure}[htbp]
  \centering
  \ContinuedFloat   
    
    \begin{subfigure}[b]{1.0\textwidth}
    \includegraphics[width=1\linewidth]{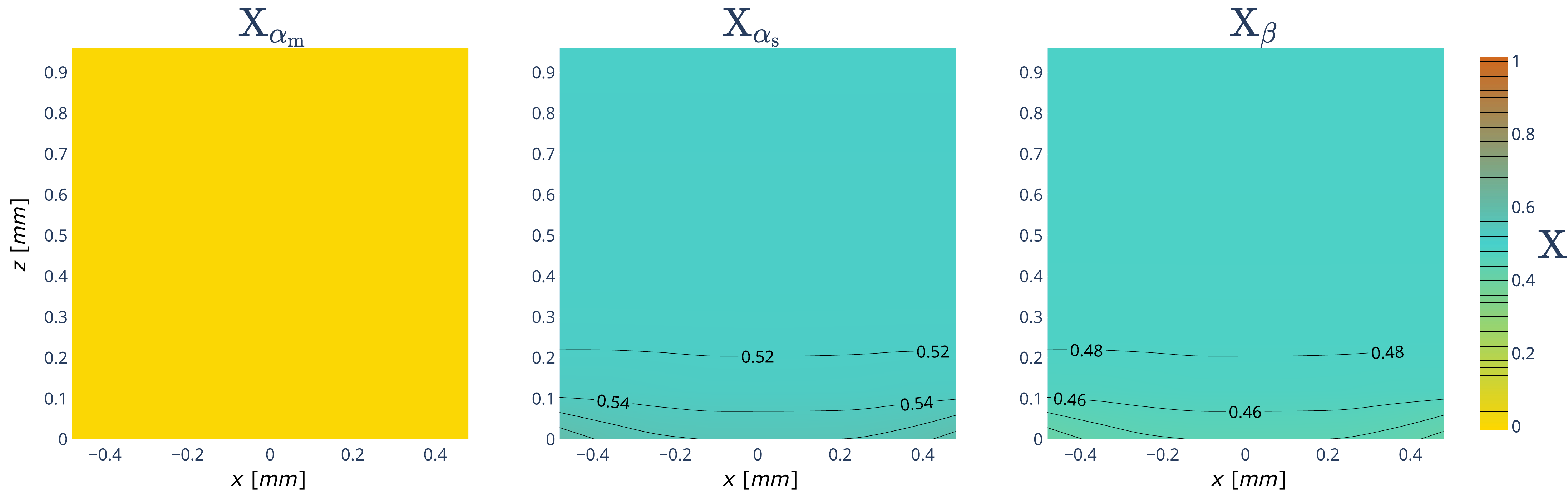}
    \newsubcap{Simulated microstructure distributions for a SLM process of a one millimeter-sided cube within its vertical x-z-center plane that contains the center node $P$. The microstructure state is shown for a base-plate temperature of $T_{\text{bp}}=900 \ K$ and the processing time $t=15 \ s$. 
    Due to the preheated base-plate no Martensite is formed. The diffusional phase transition from $\as$ to $\beta$ is not finished at the depicted snap shot.}
    \label{fig: plane_900K15s}
    \end{subfigure}
\caption{Microstructure distributions within the vertical x-z-center plane that contains the center node $P$ of the one millimeter-sided cube. The sub-figures show the resulting microstructure state of the phase-fractions $\Xa, \Xm$ and $\Xb$ at different times of the SLM process and or two different base-plate temperatures $T_{\text{bp}}=303 \ K$ and $T_{\text{bp}}=900 \ K$.}
\end{figure}   

\FloatBarrier
\section{Part-scale demonstration: Microstructure evolution during a quenching process} 
\label{sec: scaling_effects}
As SLM simulations with consistently resolved laser heat source for larger geometries (one centimeter sided cubes and above) are still hampered by the associated computational costs, even when most modern HPC systems are considered, we now want to investigate geometry-scaling effects and the effect of the increased thermal mass on the microstructure through rapid convective cooling of a 100 mm - sided cube. The numerical examples do not intend to quantitatively mimic SLM processes but rather demonstrate qualitative microstructural characteristics that evolve for parts of  practical relevant size under rapid cooling or quenching. These kind of heat treatments have broad applications in metal processing. 

\begin{figure}[htbp]
  \centering
    \includegraphics[scale=3.7]{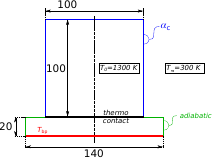}
    \caption{Schematic set-up for cooling simulation of a 100 mm - sided cube on a base-plate. The initial temperature of the cube is prescribed to $T_0=1300\ K$. The atmospheric temperature is set to $T_\infty=300 \ K$ and the free surfaces of the cube have a convective boundary condition (blue) with heat transfer coefficient $\alpha_{\text{c}}=1000\ \frac{W}{m^2 K}$. The base-plate has a Dirichlet boundary condition on the bottom (red) of first $T_{\text{bp}}=300\ K$ and, in a second simulation run, of $T_{\text{bp}}=900\ K$ and is assumed adiabatic on its free surfaces (green). The cube and the base-plate are modelled to be in thermo-mechanical contact (bold black line) using a thermal contact resistance that is equivalent to an effective heat-transfer coefficient of $\alpha_{\text{tc}}=5\cdot10^5 \frac{W}{m^2 K}$.}
    \label{fig: cube_schematic}
 \end{figure}

We model a preheated cube at $T_0=1300 \ K$ which is in thermo-mechanical contact with a base-plate with a Dirichlet boundary condition at the bottom of first $T_{\text{bp}}=300\ K$ and then in a second investigation of $T_{\text{bp}}=900\ K$ (see Figure \ref{fig: cube_schematic}, red line). The cube is subject to a convective boundary condition with heat-transfer coefficient $\alpha_{\text{c}}=1000\ \frac{W}{m^2 K}$ on its free surfaces (Figure \ref{fig: cube_schematic}, blue lines). The surrounding atmospheric temperature is set to room temperature at $T_\infty=300\ K$. The interface between base-plate and the bottom of the cube is modeled via thermo-mechanical contact interaction employing a numerical formulation recently developed in \cite{seitz2018computational} and choosing a thermal contact resistance that is equivalent to an effective heat-transfer coefficient of $\alpha_{\text{tc}}=5\cdot10^5 \frac{W}{m^2 K}$ (Figure \ref{fig: cube_schematic}, bold black line). The remaining free surfaces of the base-plate are assumed to be adiabatic (see Figure \ref{fig: cube_schematic}, green lines).
Note that the order of magnitude of the heat transfer coefficient $\alpha_{\text{c}}=1000\frac{W}{m^2 K}$ at the free surfaces of the cube was chosen to mimic convective cooling, e.g., due to forced air flow. As we only want to demonstrate qualitative results here, we omit a further detailed description of a specific cooling scenario. These quenching simulations were conducted with our in-house multi-physics framework \emph{BACI} \cite{baci}.

\begin{figure}[htbp]
  \centering
    \includegraphics[scale=0.42]{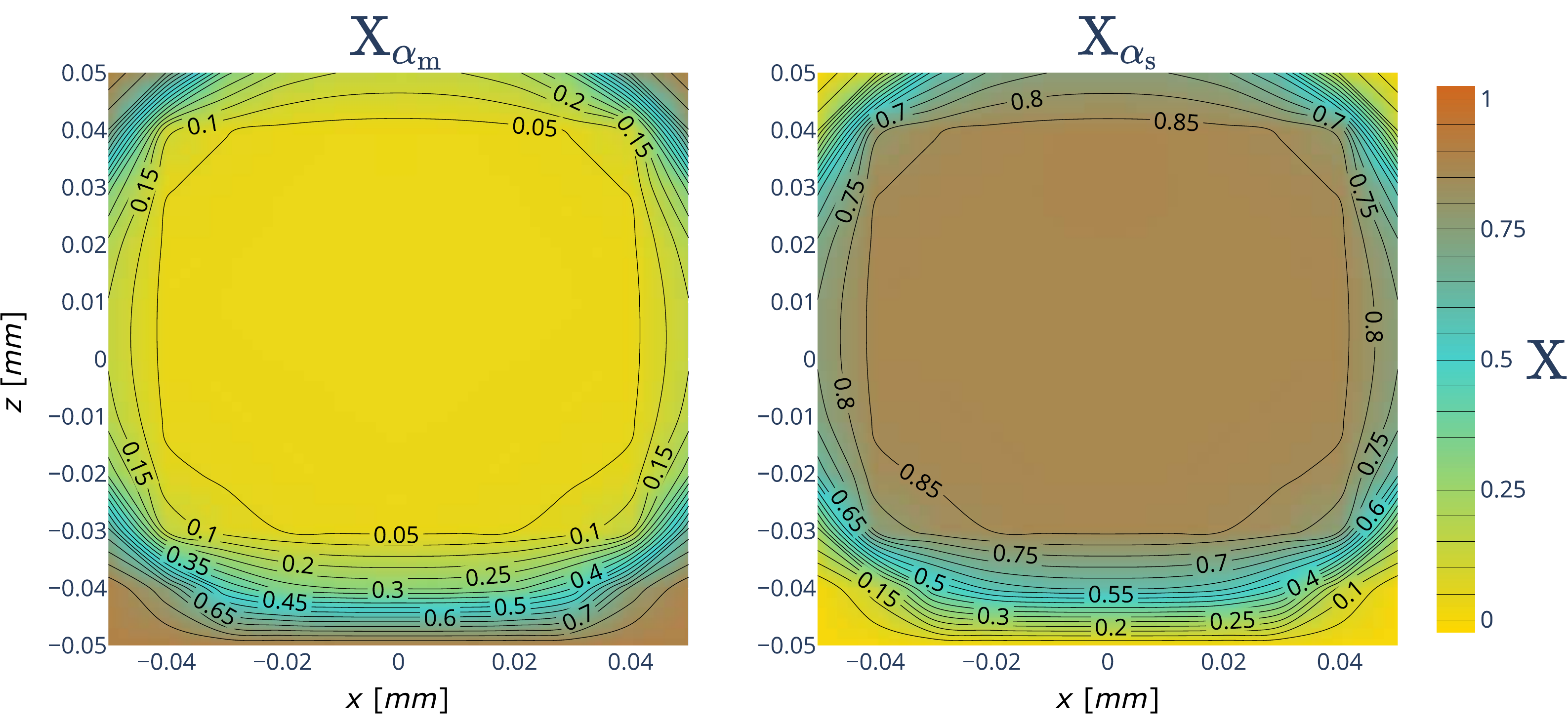}
    \caption{Simulated microstructure distribution of $\as$- and $\am$-phases in the vertical center plane of the 10 cm - sided cube after 5000 s of cooling. The simulation used a thermal heat-transfer coefficient of $\alpha_{\text{c}}=1000\ \frac{W}{m^2 K}$, a base-plate temperature of $T_{\text{bp}}=300 \ K$ and a thermal contact resistance that is equivalent to an effective heat-transfer coefficient of $\alpha_{\text{tc}}=5\cdot10^5\ \frac{W}{m^2 K}$.}
    \label{fig: cube_Tbp_300}
 \end{figure}

\begin{figure}[htbp]
  \centering
    \includegraphics[scale=0.42]{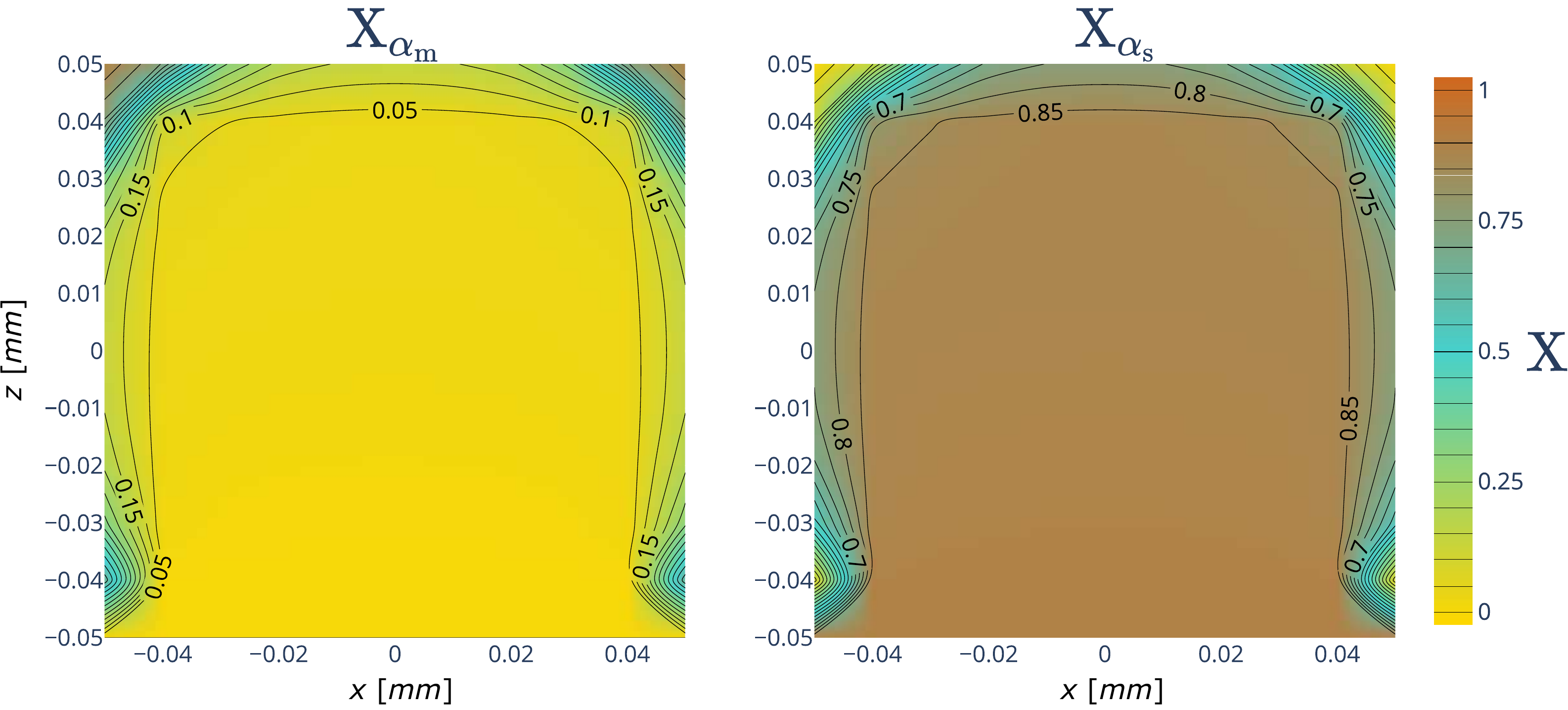}
    \caption{Simulated microstructure distribution of $\as$- and $\am$-phases in the vertical center plane of the 10 cm - sided cube after 5000 s of cooling. The simulation used a thermal heat transfer coefficient of $\alpha_{\text{c}}=1000\ \frac{W}{m^2 K}$, a base-plate temperature of $T_{\text{bp}}=900 \ K$ and a thermal contact resistance that is equivalent to an effective heat-transfer coefficient of $\alpha_{\text{tc}}=5\cdot 10^5\  \frac{W}{m^2 K}$.}
    \label{fig: cube_Tbp_900}
 \end{figure}

As a consequence of the convective cooling and the thermo-mechanical contact, the cube starts cooling down until thermodynamic equilibrium is reached. We simulated the microstructure evolution for this cooling process and investigated the resulting steady-state microstructure state at $t=5000\ s$. Figure \ref{fig: cube_Tbp_300} shows the resulting crystallographic distribution for a base-plate temperature of $T_{\text{bp}}=300 \ K$ and Figure \ref{fig: cube_Tbp_900} demonstrates the result for $T_{\text{bp}}=900 \ K$.
 
For both investigated cases, we see a several millimeter strong Martensite coating, which is especially pronounced at the corners of the cube that are characterized by higher temperature rates and lower temperature levels. These qualitative findings are in good agreement with experimental results for quenching of Ti-6Al-4V as well as additively manufactured parts \cite{Kelly.Diss, kok2016geometry}. In case of $T_{\text{bp}}=300\ K$ (Figure \ref{fig: cube_Tbp_300}) we observe an even higher Martensite amount on the base-plate-cube interface due to the higher heat transfer at the base-plate. This effect is inverted when the base-plate is heated to $T_{\text{bp}}=900 \ K$ (Figure \ref{fig: cube_Tbp_900}) and stable $\as$-phase can be found instead of the previous Martensite phase. 
Similar to the preceding SLM demonstration, a pre-heated base-plate resulted in a significant reduction of metastable Martensite phase for material close to the cube-base-plate interface. The core of the material in both demonstrations (Figure \ref{fig: cube_Tbp_300} and Figure \ref{fig: cube_Tbp_900}) is composed of stable $\as$-phase due to the increased thermal mass of the material when compared to our former SLM examples with a 1 mm sided cube in Section \ref{sec: slm_cross_sec}.

Also for larger SLM-manufactured geometries, Martensite can be expected mostly in proximity to surfaces in form of a Martensite coating, which is in compliance to experimental findings. During the SLM process, temperature rates in regions close to the melt-pool are extremely high such that we would initially expect a purely martensitic transformation as soon as the temperature decreases below the Martensite start-temperature $\Tms$. Nevertheless, the re-heating of the material during the processing of subsequent layers and tracks leads to lower cooling rates in subsequent thermal cycles on the other hand, and to an increasing overall temperature niveau within the deposited material. This effect might lead to longer dwelling times with conditions that are suitable for diffusion-based Martensite dissolution into $\as$-phase such that the core of larger SLM processed parts can also be dominated by the stable $\as$-phase, depending on the specific geometry and scanning strategy. With the development of more efficient part-scale simulation strategies for selective laser melting, we are planning on answering these questions in future investigations.
\FloatBarrier
\section{Conclusion and Outlook}
\label{sec: conclusion}
In this work, we proposed a novel physics-based and data-supported phenomenological microstructure model for part-scale simulation of Ti-6Al-4V selective laser melting (SLM). The model predicts spatially homogenized phase fractions of the most relevant microstructural species, namely the stable $\beta$-phase, the stable $\as$-phase as well as the metastable Martensite $\am$-phase, in a physically consistent manner, i.e. on the basis of pure energy and mobility competitions among the different phases. The formulation of the underlying evolution equations in rate form allows to consider general heating/cooling scenarios with temperature or time-dependent diffusion coefficients, arbitrary temperature profiles, and multiple coexisting phases in a mathematically consistent manner, which is in contrast to existing approximations via JMAK-type closed-form solutions.

Altogether, the model contains six free (physically motivated) parameters determined in a robust inverse identification process on the basis of comprehensive experimental time-temperature transformation (TTT) and transient heating data sets. Subsequently, it has been demonstrated that the identified model predicts common experimental procedures such as continuous-cooling transformation (CCT) experiments \cite{Ahmed.1998} with high accuracy and reflects well-known dynamic characteristics such as long term equilibria or critical cooling rates naturally and without the need for heuristic transformation criteria as often applied in existing models.

The part-scale simulation of selective laser melting processes with consistently resolved laser heat source is still an open research questions. In contrast to existing microstructure models that resolve the length scale of individual crystals/grains, the proposed continuum approach has the potential for such part-scale application scenarios. To the best of the authors' knowledge, in the present contribution a homogenized microstructure model of this type has for the first time been applied to predict CCT- and TTT-diagrams, which are essential means of material/microstructure characterization, and to predict the microstructure evolution for a realistic SLM application scenario (employing a state-of-the-art macroscale SLM model) and for the cooling/quenching process of a Ti-6Al-4V cube with practically relevant dimensions.

In the investigated SLM process, Martensite could be identified as the dominating microstructure species due to the process-typical extreme cooling rates and the comparatively small part size considered in the present study. A preheating of the material (e.g. via a preheated base-plate) resulted in a decreased Martensite formation and higher phase fractions of the stable $\as$-phase. In a subsequent simulation, the rapid cooling/quenching process of a larger cube (side length $10\ cm$) resting on a cold metal plate and subject to free convection on the remaining surfaces was considered. It was demonstrated that the high cooling rates in near-surface domains can lead to a strong martensitic coating of several millimeters, a behavior that is well-known from practical quenching experiments. The slower average cooling rates and the higher thermal mass in this example resulted in a large core domain dominated by the stable $\as$-phase.

In future research work, the proposed microstructure model shall be employed to inform a nonlinear elasto-plastic constitutive model, thus contributing to the long-term vision of achieving accurate thermo-mechanical simulations of selective laser melting processes on part-scale. Furthermore, SLM process parameters shall be inversely adjusted to yield specific microstructural distributions and hence desired mechanical properties by deploying novel efficient multi-fidelity approaches for (inverse) uncertainty propagation \cite{nitzler2020generalized}.

\section*{Acknowledgements}
This work was partially performed under the auspices of the U.S. Department of Energy by Lawrence Livermore National Laboratory under contract DE-AC52-07NA27344. In addition, the authors wish to acknowledge funding of this work by the Deutsche Forschungsgemeinschaft (DFG, German Research Foundation) within project 437616465. Especially, we want to acknowledge Robert M. Ferencz for his assistance and fruitful discussions. Finally, we want to thank Sebastian Pr\"oll, Abhiroop Satheesh and Nils Much for their support in model verification.

\printbibliography
\end{document}